\documentclass[twocolumn,aps,showpacs,amssymb,floatfix]{revtex4}

\usepackage{graphicx}
\usepackage{xcolor}
\usepackage{bm}
\usepackage{epsf}
\usepackage{epsfig}
\usepackage{amsmath}
\usepackage{amssymb}
\usepackage{amsfonts}
\usepackage{float}
\usepackage{hyperref}
\usepackage{dsfont}  
\usepackage{slashed}  
\usepackage{booktabs}
\usepackage[sort&compress]{natbib}

\newcommand{\be}{\begin{equation}}  
\newcommand{\ee}{\end{equation}}  
\newcommand{\beq}{\begin{eqnarray}}  
\newcommand{\eeq}{\end{eqnarray}}

\begin{document}

\title{Disconnected quark loop contributions to nucleon observables in lattice QCD}

\date{\today}
\author{A. Abdel-Rehim$^{(a)}$, C.~Alexandrou~$^{(a,b)}$, M. Constantinou~$^{(b)}$, V. Drach~$^{(c)}$, K. Hadjiyiannakou~$^{(b)}$, K.~Jansen~$^{(b,c)}$, G. Koutsou~$^{(a)}$, A. Vaquero~$^{(a)}$}
\affiliation{
  $^{(a)}$ Computation-based Science and Technology Research  
  Center, Cyprus Institute, 20 Kavafi Street, Nicosia 2121, Cyprus \\  
  $^{(b)}$ Department of Physics, University of Cyprus, P.O. Box 20537, 1678 Nicosia, Cyprus\\  
  $^{(c)}$ NIC, DESY, Platanenallee 6, D-15738 Zeuthen, Germany}

\begin{abstract}
  
  We perform a high statistics calculation of 
  disconnected fermion loops  on Graphics Processing
  Units  for a range
  of nucleon matrix elements extracted using lattice QCD.
  The isoscalar electromagnetic and axial vector form factors, the sigma terms and the  momentum fraction and
  helicity are among the quantities we evaluate.
  We compare the disconnected contributions to the connected ones and give the physical implications on nucleon observables that probe its structure. 
\end{abstract}  


\maketitle 

\setcounter{figure}{\arabic{figure}}

\newcommand{\twopt}[5]{\langle G_{#1}^{#2}(#3;\mathbf{#4};\Gamma_{#5})\rangle}
\newcommand{\threept}[7]{\langle G_{#1}^{#2}(#3,#4;\mathbf{#5},\mathbf{#6};\Gamma_{#7})\rangle}  

\newcommand{\Op}{\mathcal{O}} 
\newcommand{\C}{\mathcal{C}} 
\newcommand{\eins}{\mathds{1}} 

\bibliographystyle{apsrev}

\section{Introduction}

Lattice QCD simulations are currently performed near or at the
physical value of the light quark mass. This allows a study of hadron
structure that can provide valuable information for phenomenology and
experiment. However, a number of important observables are computed
neglecting disconnected quark loop contributions.  The evaluation of
disconnected quark loops is therefore of paramount importance if we
want to eliminate a systematic error inherent in the determination of
hadron matrix elements in lattice QCD. The computation of disconnected
quark loops within the lattice QCD formulation requires the
calculation of the so-called all-to-all or time-slice-to-all
propagators, for which the computational resources required to
estimate them with, e.g. stochastic methods, are much larger than
those required for the corresponding connected contributions. In
addition, they are prone to large gauge noise. It is for these
reasons that in most hadron structure studies up to now the
disconnected contributions were neglected, introducing an uncontrolled
systematic uncertainty~\cite{Alexandrou:2011iu}.

Recent progress in algorithms, however, combined with the increase in
computational power, have made such calculations feasible. On the
algorithmic side, a number of improvements like the one-end trick~\cite{Boucaud:2008xu, Michael:2007vn, Dinter:2012tt}, dilution~\cite{Bernardson:1993he, Viehoff:1997wi, O'Cais:2004ww, Foley:2005ac,
  Alexandrou:2012zz}, the Truncated Solver Method (TSM)~\cite{Alexandrou:2012zz, Collins:2007mh, Bali:2009hu} and the Hopping
Parameter Expansion (HPE)~\cite{Boucaud:2008xu, McNeile:2000xx} have
led to a significant reduction in both stochastic and gauge noise
associated with disconnected quark loops. Moreover, using special
properties of the twisted mass fermion Lagrangian~\cite{Frezzotti:2000nk,Frezzotti:2003xj,Frezzotti:2003ni,Frezzotti:2004wz} one can
further enhance the signal-to-noise ratio by taking the appropriate
combination of flavors. On the hardware side, graphics cards (GPGPUs
or GPUs) can provide a large speedup in the evaluation of quark
propagators and contractions. In particular, for the TSM, which
relies on a large number of inversions of the Dirac matrix in single
or half precision, GPUs provide an optimal platform.

In this paper, the aim is to use our findings on the performance of
recently developed methods~\cite{Alexandrou:2013wca} to compute to
high accuracy the disconnected contributions that enter in the
determination of nucleon form factors, sigma terms and first moments
of parton distributions. The evaluation will be performed using one
ensemble generated with two light degenerate quarks and a strange and
charm quark with masses fixed to their physical values ($N_f=2+1+1$)
using the twisted mass fermion discretization. The lattice size is
$32^3\times 64$, the lattice spacing extracted from the nucleon
mass~\cite{Alexandrou:2013joa} $a=0.082(1)(4)$~fm and the pion mass about
370~MeV. This ensemble will be hereafter referred to as the B55.32
ensemble. The aim is to compare the disconnected contributions
computed using ${\cal O}(10^5)$ measurements to the connected ones and
assess the importance of the disconnected contributions to nucleon
observables computed in lattice QCD for this given ensemble.  The
paper is organized as follows: in Section~\ref{sec:disc_methods} we
summarize the algorithms and variance reduction techniques
employed, and in Section~\ref{sec:fullstats} we present the main
numerical results of this paper, namely the disconnected
contributions to nucleon generalized form factors. In Section~\ref{sec:connected} we compare the
disconnected contributions with the corresponding connected ones. In
Section~\ref{sec:conclusions} we give our conclusions and outlook.

\section{Methods for disconnected calculations}
\label{sec:disc_methods}
\subsection{Truncated Solver Method \label{stoSec}}
The exact computation of all-to-all (or time-slice-to-all) propagators on
a lattice volume of physical interest is outside our current computer
power, since this would require volume (or spatial volume) times
inversions of the Dirac matrix, whose size ranges from $\sim 10^7$ for
a $24^3\times 48$ lattice to $\sim 10^9$ for the largest volumes of
$96^3\times 192$ considered nowadays. We will use the Truncated Solver
Method (TSM) combined with the one-end trick to evaluate the disconnected
contributions. This method was shown to be optimal for most
observables involved in nucleon structure computations~\cite{Alexandrou:2013wca}. For completeness we summarize here the
methods and refer the reader to Ref.~\cite{Alexandrou:2013wca} for a
more detailed description and the comparison against other methods.

The usual approach to evaluate disconnected contributions is to
compute an unbiased stochastic estimate of the all-to-all propagator
\cite{Bitar:1989dn} by generating a set of $N_r$ sources
$\left|\eta_r\right\rangle$ randomly drawn from e.g.
$\mathbb{Z}_2\otimes i\mathbb{Z}_2$. Solving for
$\left|s_r\right\rangle$ in
\begin{equation}
  M\left|s_r\right\rangle = \left|\eta_r\right\rangle
  \label{etaToS}
\end{equation}
and calculating
\begin{equation}
  M_E^{-1}:=\frac{1}{N_r}\sum_{r=1}^{N_r}\left|s_r\right\rangle\left\langle\eta_r\right|\approx M^{-1}
  \label{estiM}
\end{equation}
provides an unbiased estimate of the all-to-all propagator as
$N_r\rightarrow \infty$.  Since, in general, the number of noise
vectors $N_r$ required is much smaller than the lattice volume $V$,
the computation becomes feasible. How large $N_r$ should be depends on
the observable.

\begin{figure}[h!]
  \begin{center}
    \includegraphics[width=0.4\textwidth,angle=0]{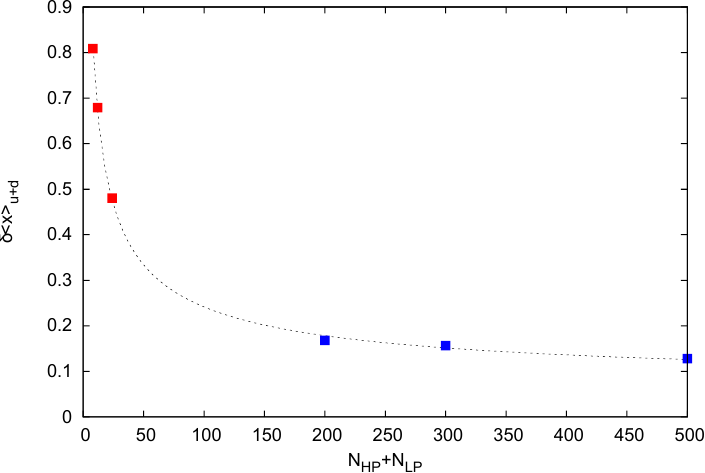}
    \caption{The error on the isoscalar momentum fraction
      $\delta\langle x \rangle_{u+d}$ as a function of $N_{\rm
        HP}+N_{\rm LP}$ for 68000 measurements. The three leftmost
      points (red squares) correspond to $N_{\rm LP}=0$ and the three
      rightmost to $N_{\rm HP}=24$. The dotted line is the result of fitting to the
      Ansatz $1/\sqrt{a+\frac{b}{N_{\rm HP}+N_{\rm LP}}}$.}
    \label{fig:x_tuning}
  \end{center}
\end{figure}

The TSM is a way to increase $N_r$ at a reduced computational
cost. The idea behind the method is the following: instead of
inverting to high precision the stochastic sources in
Eq. \eqref{etaToS}, we can aim at a low precision (LP)
estimate 
\begin{equation} \left|s_r\right\rangle_{LP} =
  \left(M^{-1}\right)_{LP}\left|\eta_r\right\rangle,
  \label{etaToSLP}
\end{equation}
where the number of inversions of the Conjugate Gradient (CG) used 
 is truncated. The criterion for the low precision inversions
can be selected by specifying a relaxed stopping condition in the CG
e.g. by allowing a relatively large value of the residual, which
in turn determines the number of iterations required to invert a source
to low precision. Following Refs.~\cite{Alexandrou:2012zz,
  Alexandrou:2013wca}, we choose a stopping condition at fixed value
of the residual $|\hat{r}|_{\rm LP}\sim 10^{-2}$. $N_{\rm HP}$ is then
selected by requiring that the bias introduced when using $N_{\rm LP}$
low precision vectors is corrected. We estimate the correction $C_E$
to the bias stochastically by inverting a number of sources to high
and low precision, and calculating the difference,
\begin{equation}
  C_E:=\frac{1}{N_{\rm HP}}\sum_{r=1}^{N_{\rm HP}}\left[\left|s_r\right\rangle_{\rm HP} - \left|s_r\right\rangle_{\rm LP}\right]\left\langle\eta_r\right|,
  \label{estiC}
\end{equation}
where the $\left|s_r\right\rangle_{\rm HP}$ are calculated by solving
Eq.~\eqref{etaToS} up to high precision, so our final estimate becomes

\begin{eqnarray}
  M_{E_{TSM}}^{-1}:=& &\frac{1}{N_{\rm HP}}\sum_{r=1}^{N_{\rm HP}}\left[\left|s_r\right\rangle_{\rm HP} - \left|s_r\right\rangle_{\rm LP}\right]\left\langle\eta_r\right| \nonumber \\
  &+&
  \frac{1}{N_{\rm LP}}\sum_{j=N_{\rm HP}}^{N_{\rm HP}+N_{\rm LP}}\left|s_r\right\rangle_{\rm LP}\left\langle\eta_r\right|,
  \label{estiTSM}
\end{eqnarray}
which requires $N_{\rm HP}$ high precision (HP) inversions and $N_{\rm
  HP}+N_{\rm LP}$ low precision inversions. The ratio of the number of
HP inversions to the LP  ones is determined with the criterion of
choosing as large a ratio as possible while still ensuring that the
final result is unbiased. In this work, we will compute fermion loops
with the complete set of $\Gamma$-matrices up to one-derivative
operators. The tuning is, thus, performed using an operator that
requires a large number of stochastic noise vectors, such as the
nucleon isoscalar momentum fraction $\langle x\rangle_{u+d}$ and we
optimize $N_{\rm HP}$ and $N_{\rm LP}$ so as to get the smallest
error at the lowest computational cost. In Fig.~\ref{fig:x_tuning} we
show the error on $\langle x\rangle_{u+d}$ as one varies $N_{\rm HP}$
and $N_{\rm LP}$. As can be seen, the error decreases like
$1/\sqrt{a+\frac{b}{N_{\rm HP}+N_{\rm LP}}}$ with $a$ and $b$ positive
parameters. Fixing $N_{\rm HP}=24$ and increasing $N_{\rm LP}$ reduces
the error rapidly until $N_{\rm LP}$ reaches about
$N_{\rm LP}\sim~300$. In Ref.~\cite{Alexandrou:2013wca} we showed that
a ratio of $N_{\rm LP}$ to $N_{\rm HP}$ of about 20 can be considered
sufficient to produce an unbiased estimate for the class of
observables considered here.  Therefore, in this work we take $N_{\rm
  HP}=24$ and choose $N_{\rm LP}=500$ for the light quark sector.  For
the strange and charm quarks we take $N_{\rm LP}=300$.  These values were
shown to also be optimal for the isoscalar axial
charge~\cite{Alexandrou:2013wca}.

\subsection{The one-end trick}
The twisted mass fermion (TMF) formulation allows the use of a very
powerful method to reduce the variance of the stochastic estimate of
the disconnected diagrams. From the discussion given in
section~\ref{stoSec}, the standard way to proceed with the computation
of disconnected diagrams would be to generate $N_r$ stochastic sources
$\eta_r$, invert them as indicated in Eq.~\eqref{etaToS}, and compute
the disconnected diagram corresponding to an operator $X$ as

\begin{eqnarray}
  \frac{1}{N_r}\sum_{r=1}^{N_r} \left\langle \eta^\dagger_r X s_r\right\rangle &=& \textrm{Tr}\left(M^{-1}X\right) \nonumber \\ & +& O\left(\frac{1}{\sqrt{N_r}}\right),
  \label{loopSt}
\end{eqnarray}
where the operator $X$ is expressed in the twisted basis. However, if
the operator $X$ involves a $\tau_3$ acting in flavor space, one can
utilize the following identity of the twisted mass Dirac operator with
$+\mu$ denoted by $M_u$ and $-\mu$ denoted by $M_d$:

\begin{equation}
  M_u - M_d = 2i\mu a\gamma_5.
\end{equation}
Inverting this equation we obtain

\begin{equation}
  M^{-1}_u - M^{-1}_d = -2i\mu aM_d^{-1}\gamma_5M_u^{-1}.
  \label{vvTrick}
\end{equation}
Therefore, instead of using Eq. \eqref{loopSt} for the operator $X\tau_3$, we can alternatively write

\begin{align}
  \frac{2i\mu a}{N_r}\sum_{r=1}^{N_r} \left\langle s^\dagger_r \gamma_5 X s_r\right\rangle =& \nonumber\\
  \textrm{Tr}\left(M_u^{-1}X\right) - \textrm{Tr}\left(M_d^{-1}X\right) &+ O\left(\frac{1}{\sqrt{N_r}}\right)=\nonumber\\
  -2i\mu a\textrm{Tr}\left(M_d^{-1}\gamma_5M_u^{-1}X\right) &+ O\left(\frac{1}{\sqrt{N_r}}\right).
  \label{loopVv}
\end{align}

Two main advantages result due to this substitution: i) the
fluctuations are effectively reduced by the $\mu$ factor, which is
small in current simulations, and ii) an implicit sum of $V$ terms
appears in the right hand side (rhs) of Eq.~\eqref{vvTrick}. The trace
of the left hand side (lhs) of the same equation develops a
signal-to-noise ratio of $1/\sqrt{V}$, but thanks to this implicit
sum, the signal-to-noise ratio of the rhs becomes $V/\sqrt{V^2}$. In
fact, using the one-end trick yields for the same operator a large
reduction in the errors for the same computational cost as compared to
not using it~\cite{Boucaud:2008xu, Michael:2007vn, Dinter:2012tt}.
A similar approach proved to be very successful in the determination of the 
$\eta^\prime$ mass~\cite{Jansen:2008wv,Ottnad:2012fv,Michael:2013gka}. 
The identity given in Eq.~\eqref{vvTrick} can only be applied when a
$\tau_3$ flavor matrix appears in the operator expressed in the
twisted basis. For other operators one can use the identity

\begin{equation}
  M_u + M_d = 2D_W,
\end{equation}
where $D_W$ is the Dirac-Wilson operator without a twisted mass term. After some algebra, one finds

\begin{eqnarray}
  \frac{2}{N_r}\sum_{r=1}^{N_r} \left\langle s^\dagger_r \gamma_5 X\gamma_5 D_W s_r\right\rangle &=& \textrm{Tr}\left(M_u^{-1}X\right) + \textrm{Tr}\left(M_d^{-1}X\right) \nonumber \\ & +& O\left(\frac{1}{\sqrt{N_r}}\right).
  \label{loopStD}
\end{eqnarray}
This lacks the $\mu$-suppression factor, which, as we will see in the following sections and as
discussed in more detail in Ref.~\cite{Alexandrou:2013wca}, introduces
a considerable penalty in the signal-to-noise ratio.

Because of the volume sum that appears in Eq.~\eqref{vvTrick} and
Eq.~\eqref{loopStD}, the sources must have entries on all sites, which
in turn means that we can compute the fermion loop at all time slices where the
operator is inserted in a single inversion. This allows us to evaluate the three-point function for
all combinations of source-sink time separation and insertion time slices,
which will prove essential in identifying the contribution of excited
state effects for the different operators.

\section{Results}
\label{sec:fullstats}

In this section we present results from a high statistics evaluation
of all the disconnected contributions involved in the evaluation of
nucleon form factors and first moments of generalized parton
distributions as well as sigma terms. As already mentioned, the
analysis is performed using an ensemble of $N_f=2+1+1$ twisted mass
configurations simulated with pion mass of $ am_\pi=0.15518(21)(33)$
and strange and charm quark masses fixed to approximately their
physical values (B55.32 ensemble)~\cite{Baron:2010bv}.  The lattice
size is $32^3\times 64$ giving $m_{\pi}L\sim 5$.  We use the one-end
trick method combined with the TSM with $N_{\rm HP}=24$ and $N_{\rm
  LP}=500$ noise vectors for the light quark loops. For the strange
and charm quark sector we use $N_{\rm HP}=24$ and $N_{\rm LP}=300$.
Using 2,300 gauge-field configurations, with 16 source positions for
the two-point function and by averaging results for the proton/neutron
and forward/backward propagating nucleons we effectively have $\sim
150,000$ measurements.

An advantage of the one-end trick is that, having the loop at all
time slices, we can combine with two-point functions produced at any
source time slice.  Furthermore, since the noise sources are defined
on all sites, we obtain the fermion loops at all insertion
time slices. We can thus compute all possible combinations of
source-sink time separations and insertion times in the three-point
function. This feature enables us to use the summation method, in
addition to the plateau method, with no extra computational effort.

The summation method has been known for a long
time~\cite{Maiani:1987by, Gusken:1999te} and has been revisited in the
study of $g_A$~\cite{Capitani:2010sg}. In both the plateau and
summation approaches, one constructs ratios of three- to two-point
functions in order to cancel unknown overlaps and exponentials in the
leading contribution such that the matrix element of the ground state
is isolated.  For general momentum transfer we consider the ratio
\begin{widetext}
\be
R(t_{\rm ins},t_s){=} \frac{G^{3pt}(\Gamma^\nu,{\vec p},{\vec q},t_{\rm ins},t_s) }{G^{2pt}(\vec{p}^\prime,
  t_s)} \hspace{-0.2cm}\sqrt{\frac{G^{2pt}(\vec p, t_s{-}t_{\rm ins})G^{2pt}(\vec p^\prime,  t_{\rm ins})G^{2pt}(\vec p^\prime,
    t_s)}{G^{2pt}(\vec p^\prime  , t_s{-}t_{\rm ins})G^{2pt}(\vec p,t_{\rm ins})G^{2pt}(\vec p,t_s)}}
\label{ratio q}
\ee
where the two- and three-point functions are given respectively  by
\beq
G^{2pt}(\vec q, t_s)=
&&\sum_{\vec x_s} \, e^{-i x_s \cdot \vec q}\,
     {\Gamma^0_{\beta\alpha}}\, \langle {J_{\alpha}(t_s,\vec x_s)}{\overline{J}_{\beta}(0,\vec{0})} \rangle \label{twop}\\
G^{3pt}(\Gamma^\nu,\vec p,\vec q,t_{\rm ins},t_s) =&&\sum_{\vec{x}_{\rm ins}, \vec x_s}\,e^{i\vec x_{\rm ins} \cdot \vec q}\, e^{-i\vec x_s \cdot \vec p}\,  \Gamma^\nu_{\beta\alpha}\, \langle
{J_{\alpha}(t_s,\vec x_s)} \Op^{\mu_1 \cdots \mu_n}(t_{\rm ins},\vec x_{\rm ins}) {\overline{J}_{\beta}(0,\vec{0})}\rangle\,.\label{threep}
\eeq
\end{widetext}
 $q=p^\prime - p$ is the momentum transfer, $t_s$ is the time
separation  between the sink and the source with the source taken at zero, and $t_{\rm ins}$ the
time separation between the current insertion and the source.  We
consider the complete set of operators  $\Op^{\mu_1,\cdots,  \mu_n}$ up
to one derivative, namely the scalar $\bar \psi\,\psi$, vector $\bar
\psi\,\gamma^{\mu}\psi$, axial-vector $\bar
\psi\,\gamma^5\,\gamma^{\mu}\psi$ and the tensor $\bar\psi
\sigma^{\mu\nu}\psi$ currents, and the one-derivative vector $\bar
\psi\,\gamma^{\{\mu_1} D^{\mu_2\}}\psi$ and  axial-vector
$\bar \psi\,\gamma_5\,\gamma^{\{\mu_1} D^{\mu_2\}}\psi$ 
 operators.  We consider kinematics for which the final momentum
$\vec p^\prime=0$ when we consider the
connected contributions. For the evaluation of disconnected contributions we
use kinematics where $\vec p= \vec{p}^\prime\ne 0$ as well as  $\vec{p}^\prime=0$.  The projection
matrices ${\Gamma^0}$ and ${\Gamma^k}$ are given by: 
\be 
{\Gamma^0} =\frac{1}{4}(\eins + \gamma_0)\,,\quad {\Gamma^k} = {\Gamma^0} i \gamma_5 \sum_{k=1}^3 \gamma_k \, .\label{proj} 
\ee 
For zero momentum transfer the ratio simplifies to 
\be
R(t_{ins},t_s)=\frac{G^{3pt}(\Gamma^\nu,\vec p,
  t_{ins},t_s)}{G^{2pt}(t_,\vec p)}
\label{ratio}
\ee
 The leading time dependence of the ratio $R(t_{\rm ins},t_s)$ is given by
\begin{equation}
  R(t_{ins},t_s) = R_{GS} + O(e^{-\Delta E_K t_{ins}}) + O(e^{-\Delta E_K(t_s-t_{ins})}),
  \label{RatioPlaDet}
\end{equation}
where $R_{GS}$ is the matrix element of interest, and the other
contributions come from the undesired excited states of energy
difference $\Delta E_K$. In the plateau method, one plots
$R(t_{ins},t_s)$ as a function of $t_{\rm ins}$.  
 For large time
separations $t_{\rm ins}$ and $t_s-t_{\rm ins}$ when excited state effects are negligible this ratio becomes a
constant (plateau region) 
and therefore fitting it to a constant yields $R_{GS}$. In the alternative
summation method, one performs a sum over $t_{\rm ins}$ to obtain:

\begin{equation}
  R_{\rm sum}(t_s) = \sum_{t_{\rm ins}=0}^{t_{\rm ins}=t_s} R(t_{\rm ins},t_s) = t_sR_{GS} + a + O(e^{-\Delta E_Kt_s})
  \label{RatioSumDet}
\end{equation}
where $a $ is a constant
and the exponential contributions coming from the excited states
decay as $e^{-\Delta E_K t_s}$ as opposed to the plateau method where
excited states are suppressed like $e^{-\Delta E_K (t_s-t_{ins})}$,
with $0\le t_{\rm ins}\le t_s$. Therefore, we
expect a better suppression of the excited states for the same $t_s$.
Note that one can exclude from the summation the initial and final time
slices  $t_s$ and
$0$ without affecting the dependence on $t_s$ in
Eq.~\eqref{RatioSumDet}. The results given in this work are obtained
excluding these contact terms from the summation.  The drawback of the
summation method is that one requires knowledge of the three point
function for all insertion times and multiple sink times and one
needs to fit to a straight line with two fitting parameters instead of
one.
\begin{table}[h]
\begin{center}
  \footnotesize
\begin{tabular}{|ccccccc|}
\hline
$Z_A$ & $Z_T$ & $Z_P$ & $Z_{DV}^{\mu\mu}$ & $Z_{DV}^{\mu\ne\nu}$ & $Z_{DA}^{\mu\mu}$ &
$Z_{DA}^{\mu\ne\nu}$ \\
0.757(3)  & 0.769(1)  &  0.506(4) & 1.019(4) &1.053(11)& 1.086(3)  & 1.105(2) \\
\hline
\end{tabular}
\caption{Renormalization constants in the chiral limit at $\beta=1.95$ in the $\overline{\rm MS}$-scheme at $\mu=2$~GeV. $Z_A$, $Z_T$ and $Z_P$ are the renormalization constants for the axial-vector, tensor and scalar  currents, and $Z_{DV}$ and $Z_{DA}$ for the one-derivative vector and axial-vector operators ${\cal O}^{\mu\nu}$. The errors given are statistical.}
\label{tab:renormalization constants}
\end{center}
\end{table}

\begin{figure}[h!]
  \includegraphics[width=0.8\linewidth,angle=0]{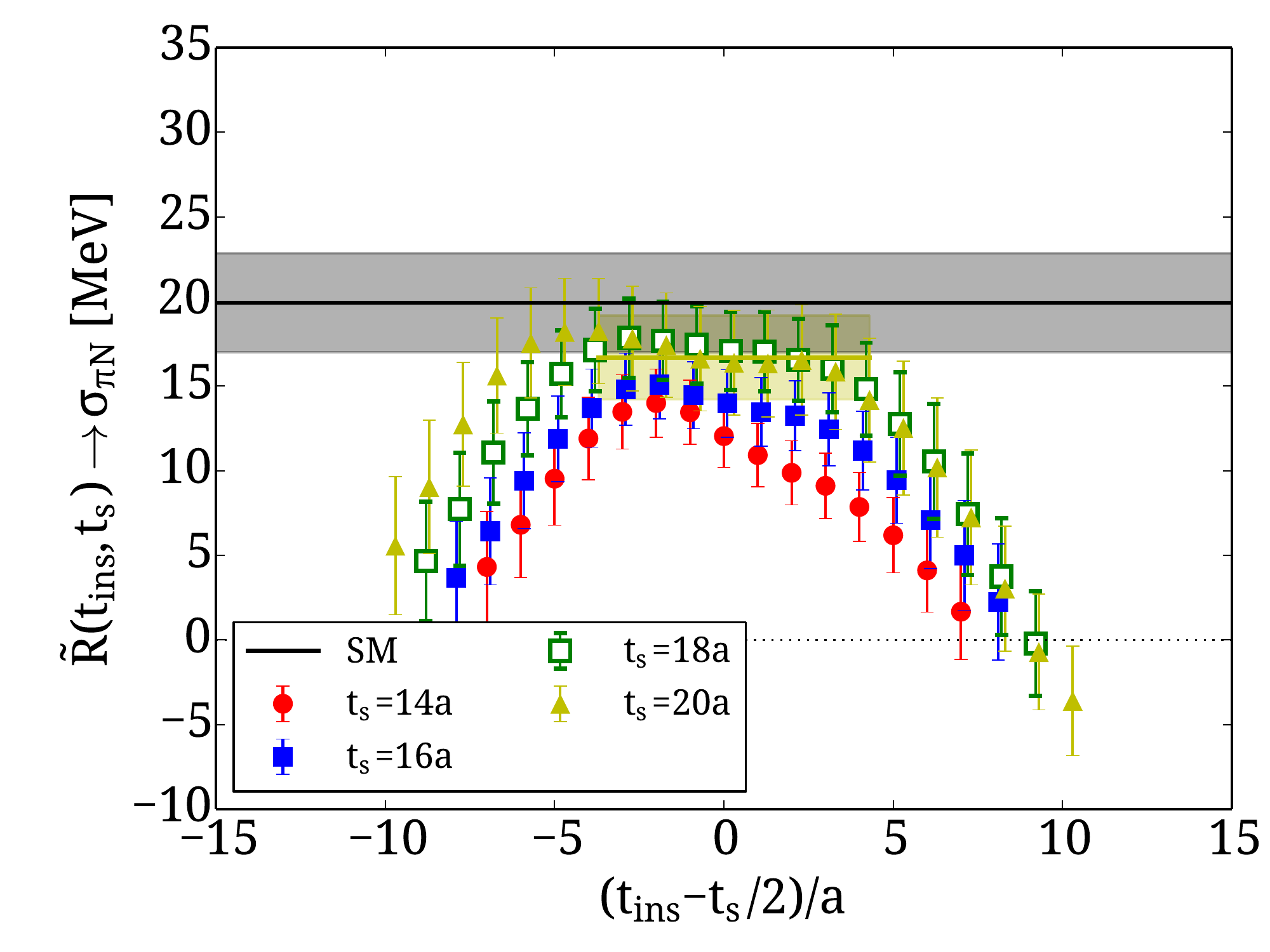}\\ 
  \includegraphics[width=0.8\linewidth,angle=0]{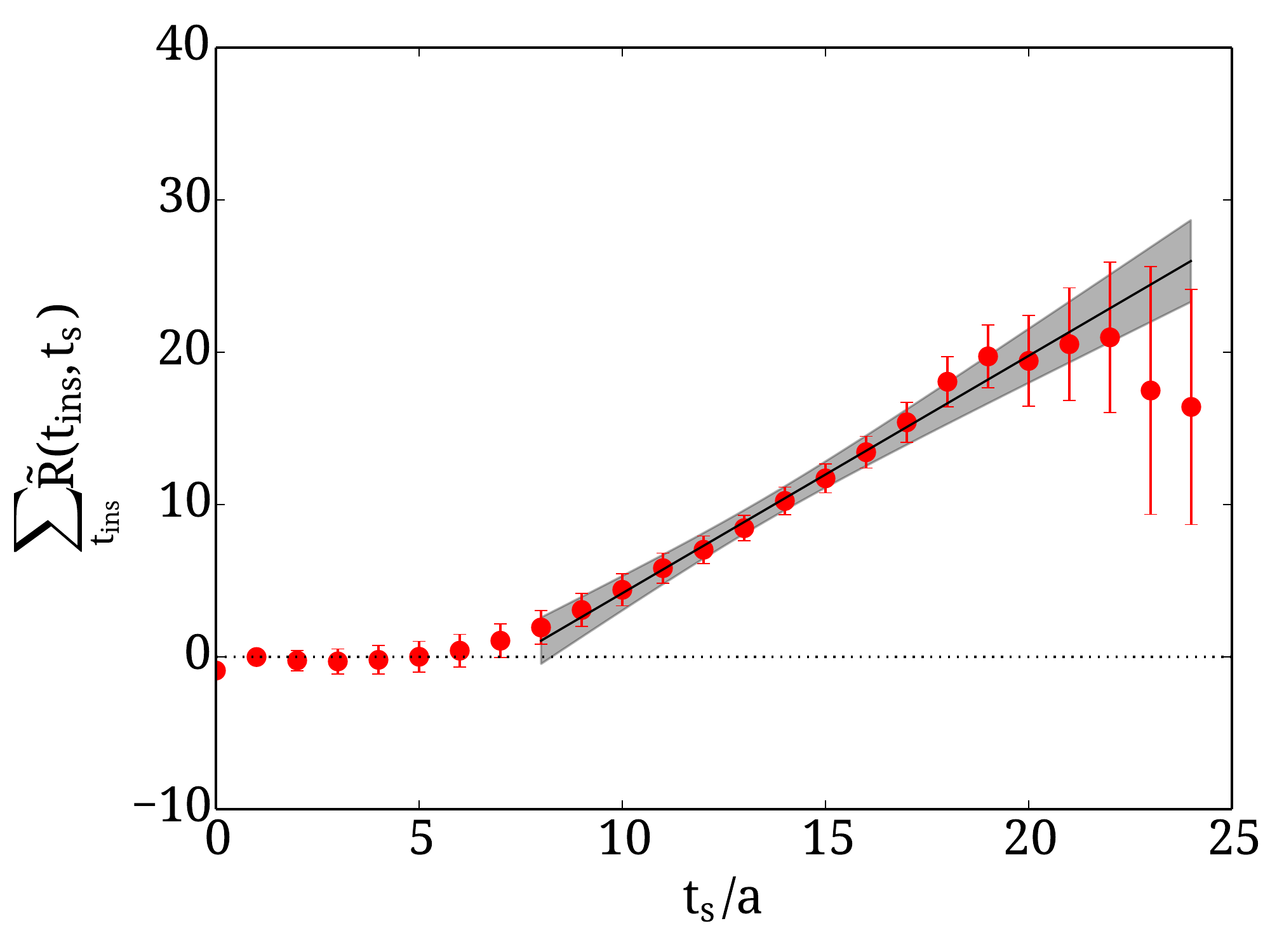}\\ 
  \includegraphics[width=0.8\linewidth,angle=0]{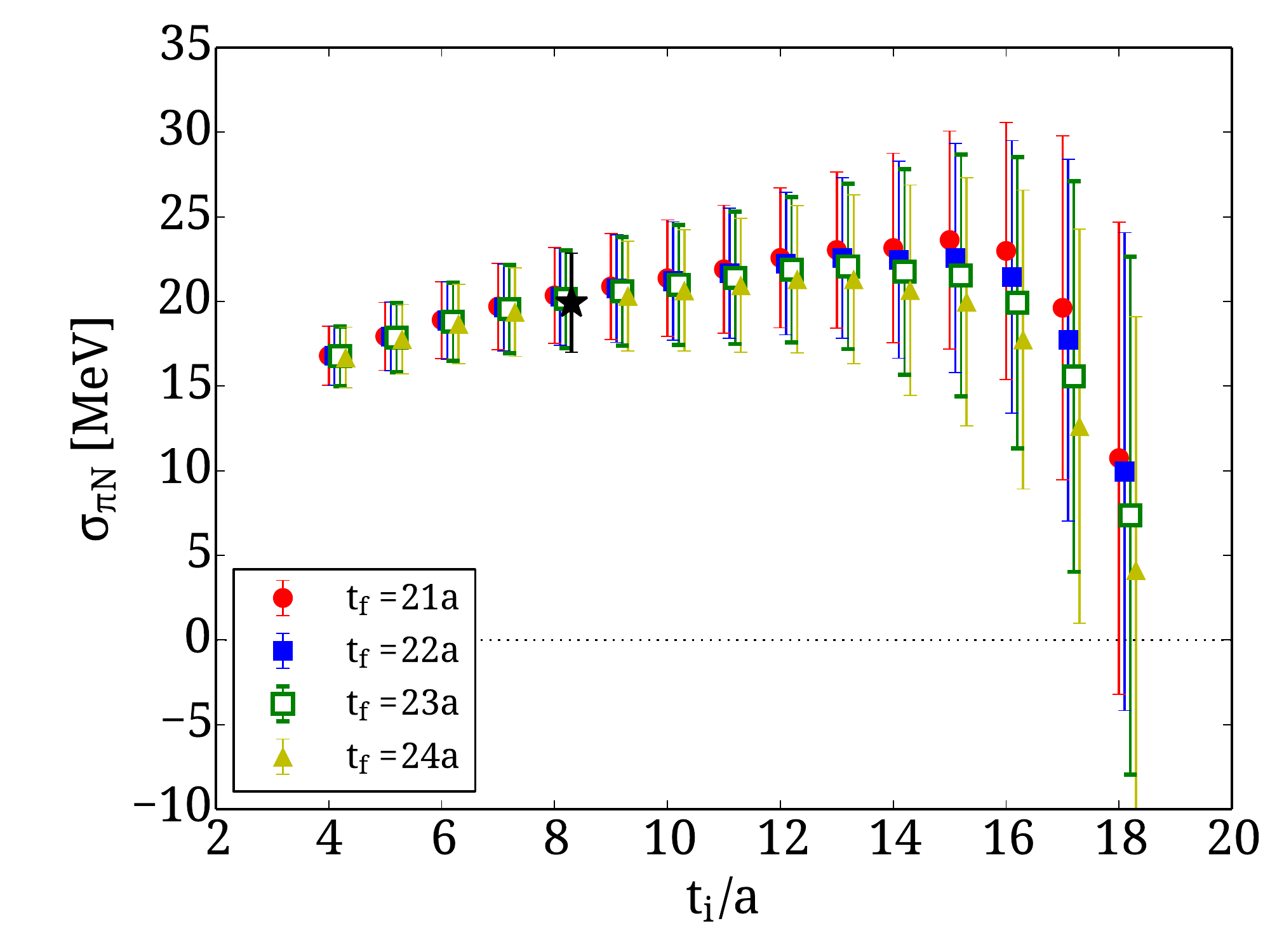}
  \caption{The disconnected contribution to the ratio from which
    $\sigma_{\pi N}$ is extracted. On the upper panel we show the
    ratio as a function of the insertion time slice with respect to
    the mid-time separation ($t_{\rm ins}-t_s/2$) for source-sink
 time separations, $t_{\rm s}=$14$a$ (red filled  circles), $t_{\rm s}=16a$
    (blue filled squares), $t_{\rm s}=18a$ (green open squares) and $t_{\rm
      s}=20a$ (yellow filled triangles).  In the central panel we show the
    summed ratio, for which the fitted slope yields the desired matrix
    element. On the lower panel we show the results obtained for the
    fitted slope of the summation method for various choices of the
    initial and final fit time slices. The star shows the
    choice for which the gray bands are plotted in the upper and
    central panels.
    \label{sigmaLight}}
\end{figure}

\begin{figure}[h!]
  \includegraphics[width=0.8\linewidth,angle=0]{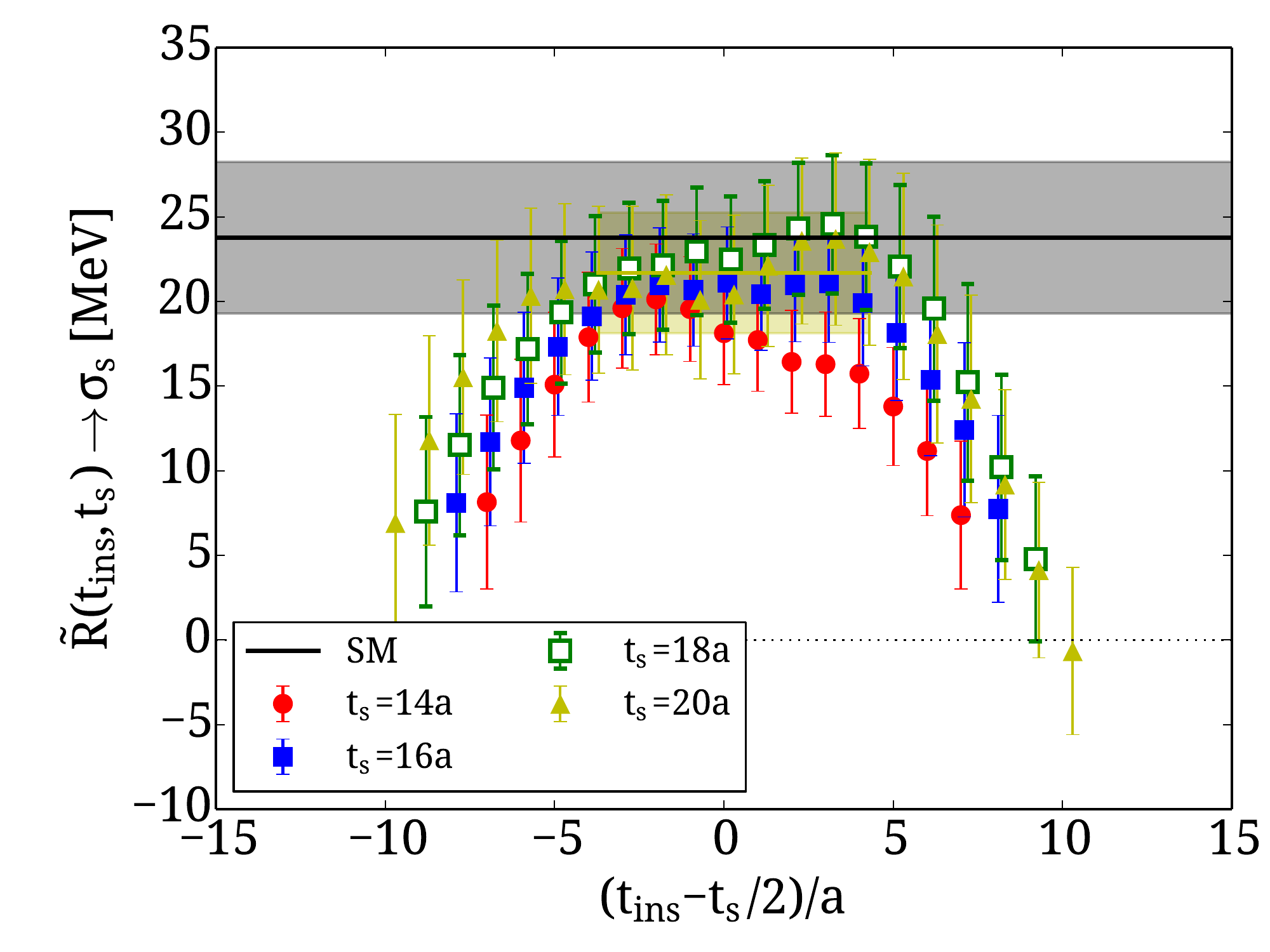}\\    
  \includegraphics[width=0.8\linewidth,angle=0]{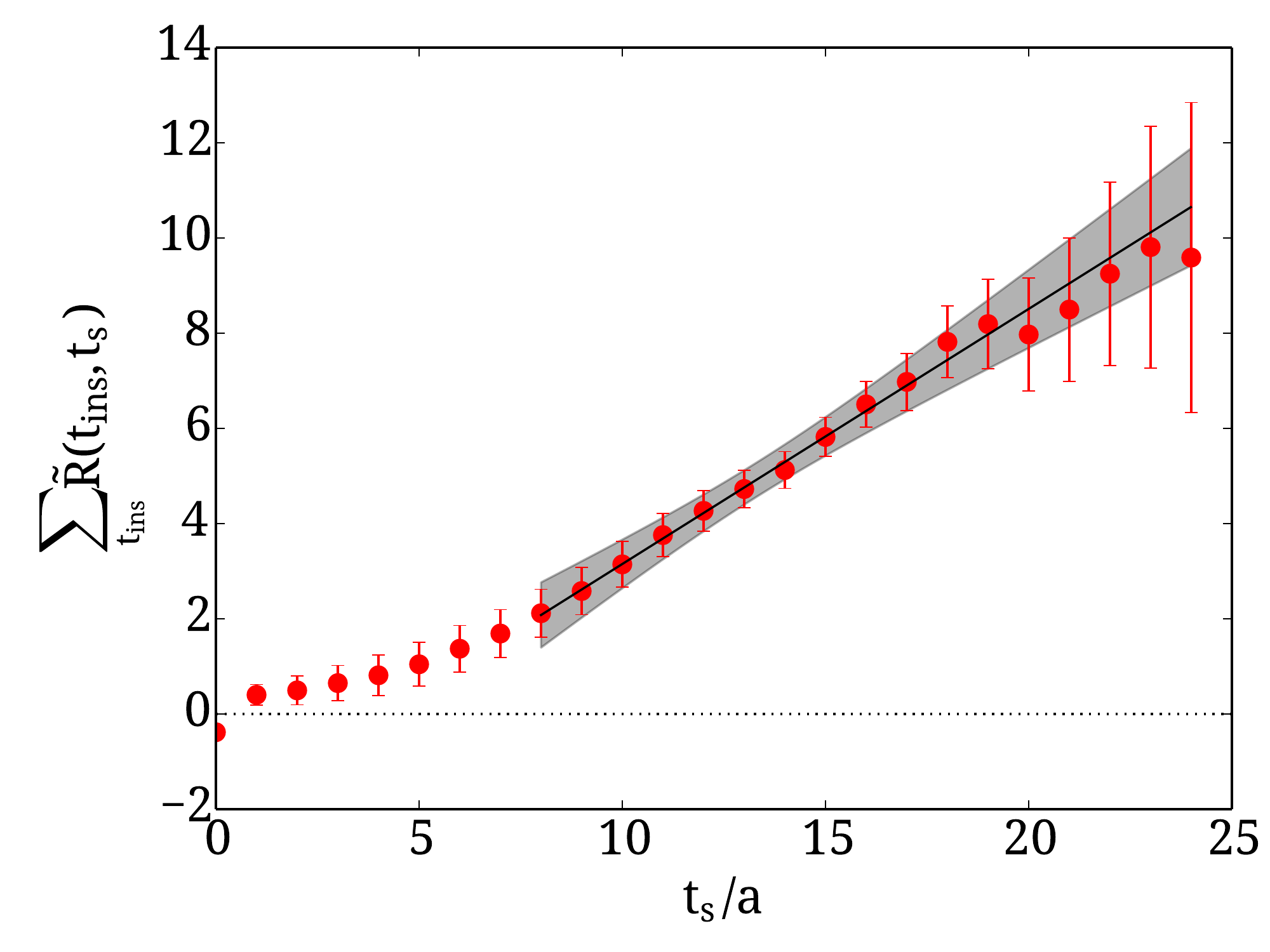}  \\       
  \includegraphics[width=0.8\linewidth,angle=0]{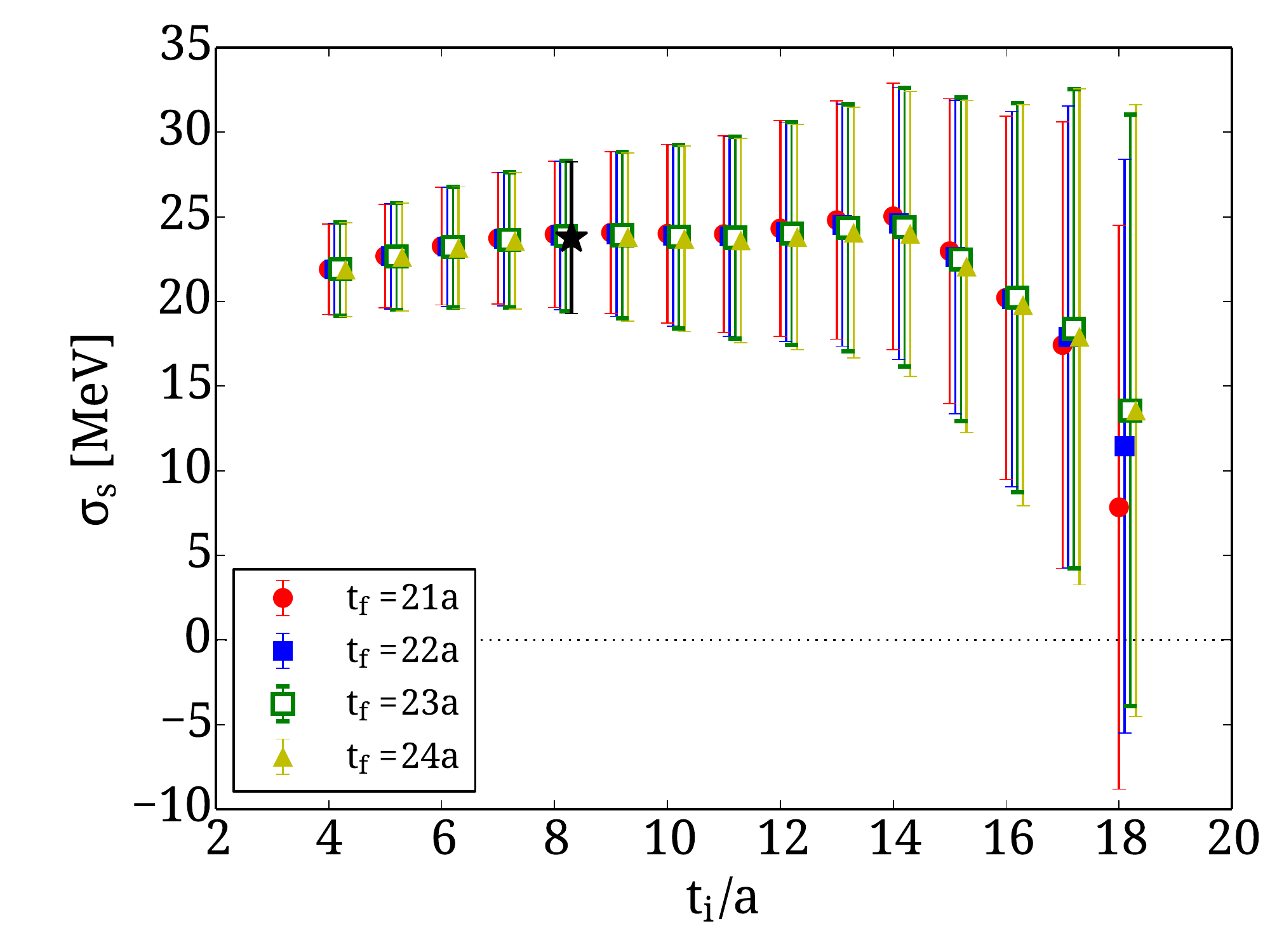} 
  \caption{The ratio from which the  strange quark content of the nucleon, 
    $\sigma_{s}$, is extracted. The notation is the same as that of
    Fig.~\ref{sigmaLight}.\label{sigmaStrange}}
\end{figure}

\begin{figure}[h!]
  \includegraphics[width=0.8\linewidth,angle=0]{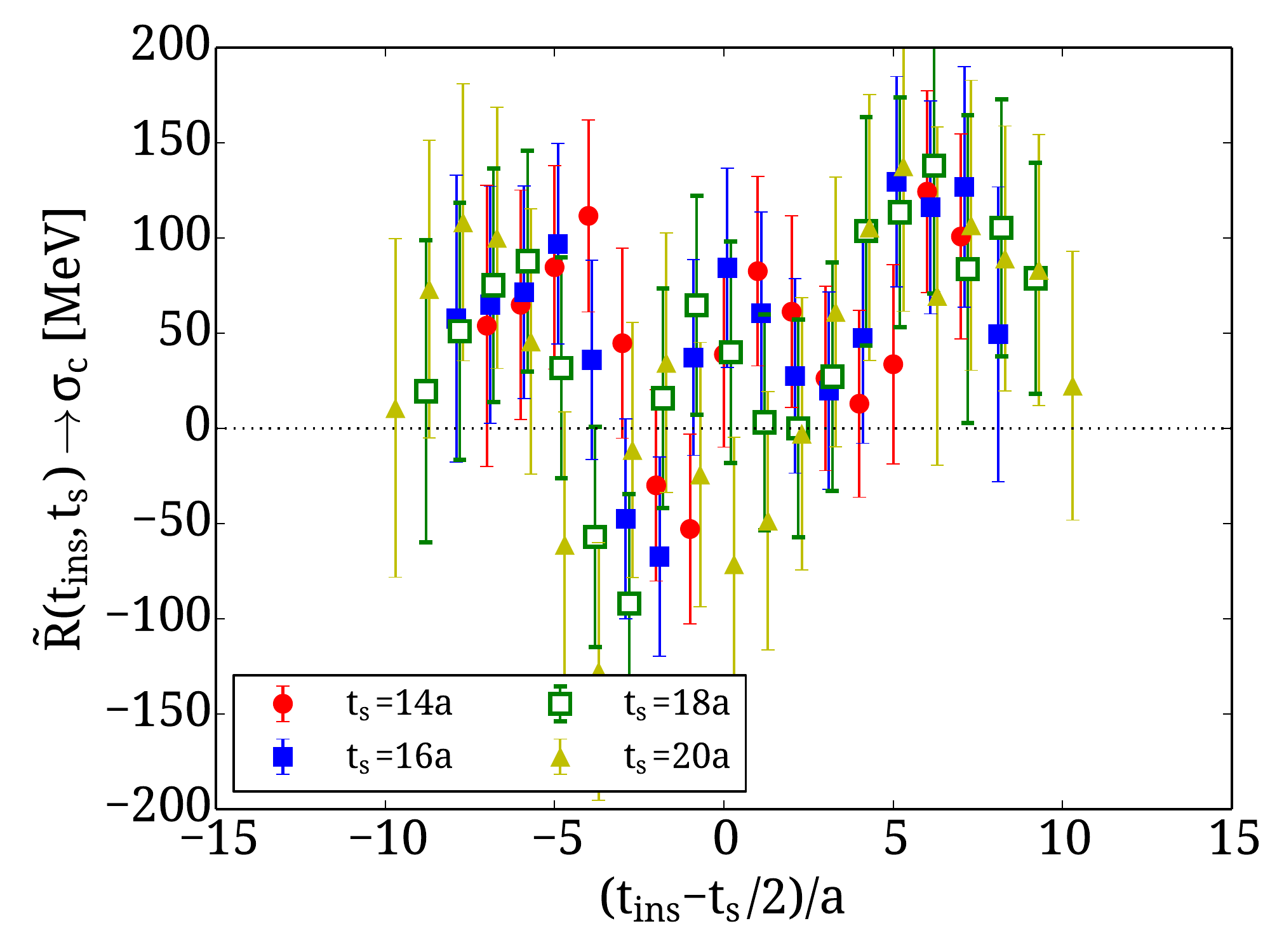} \\
  \includegraphics[width=0.8\linewidth,angle=0]{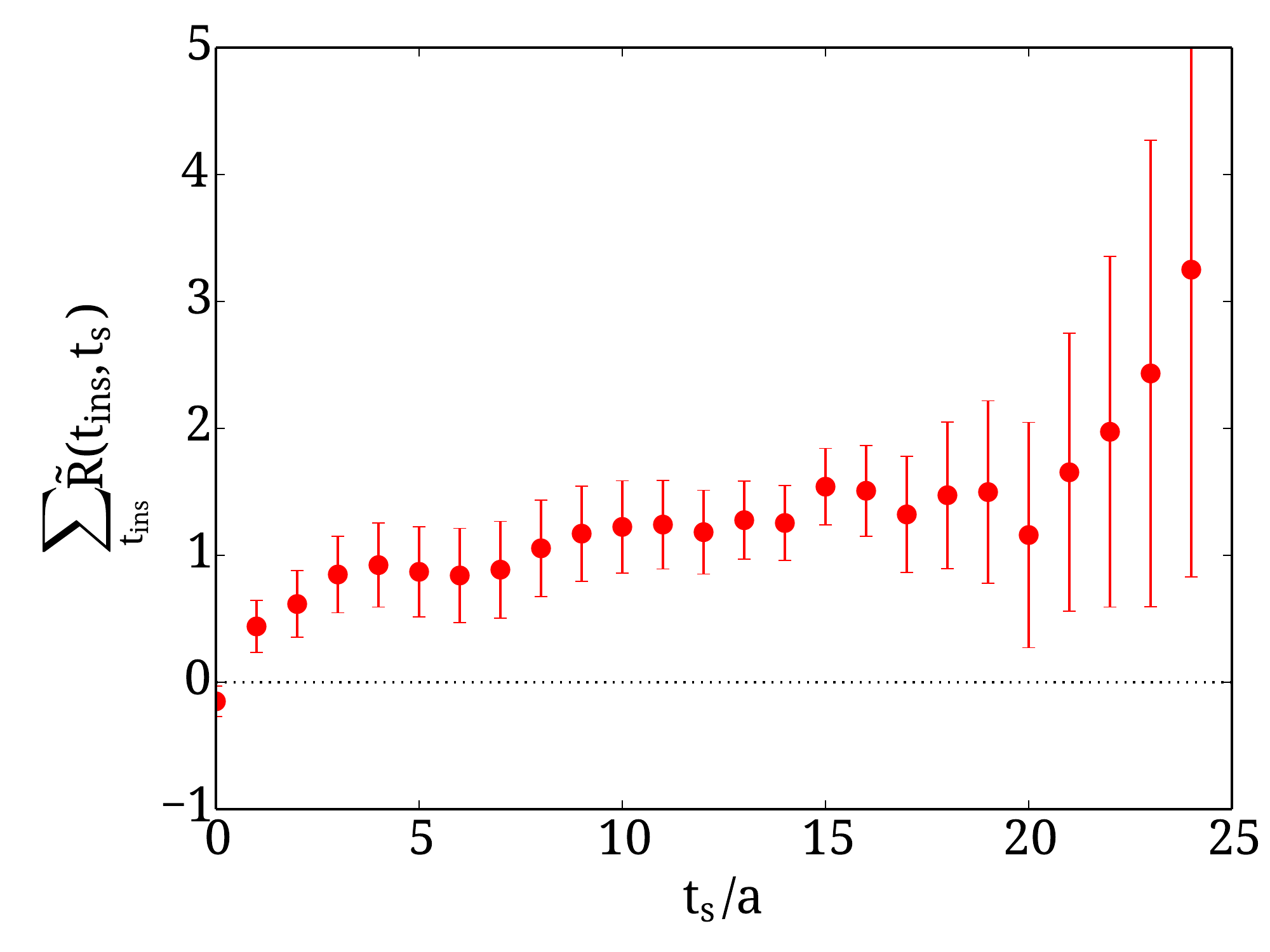} \\
  \includegraphics[width=0.8\linewidth,angle=0]{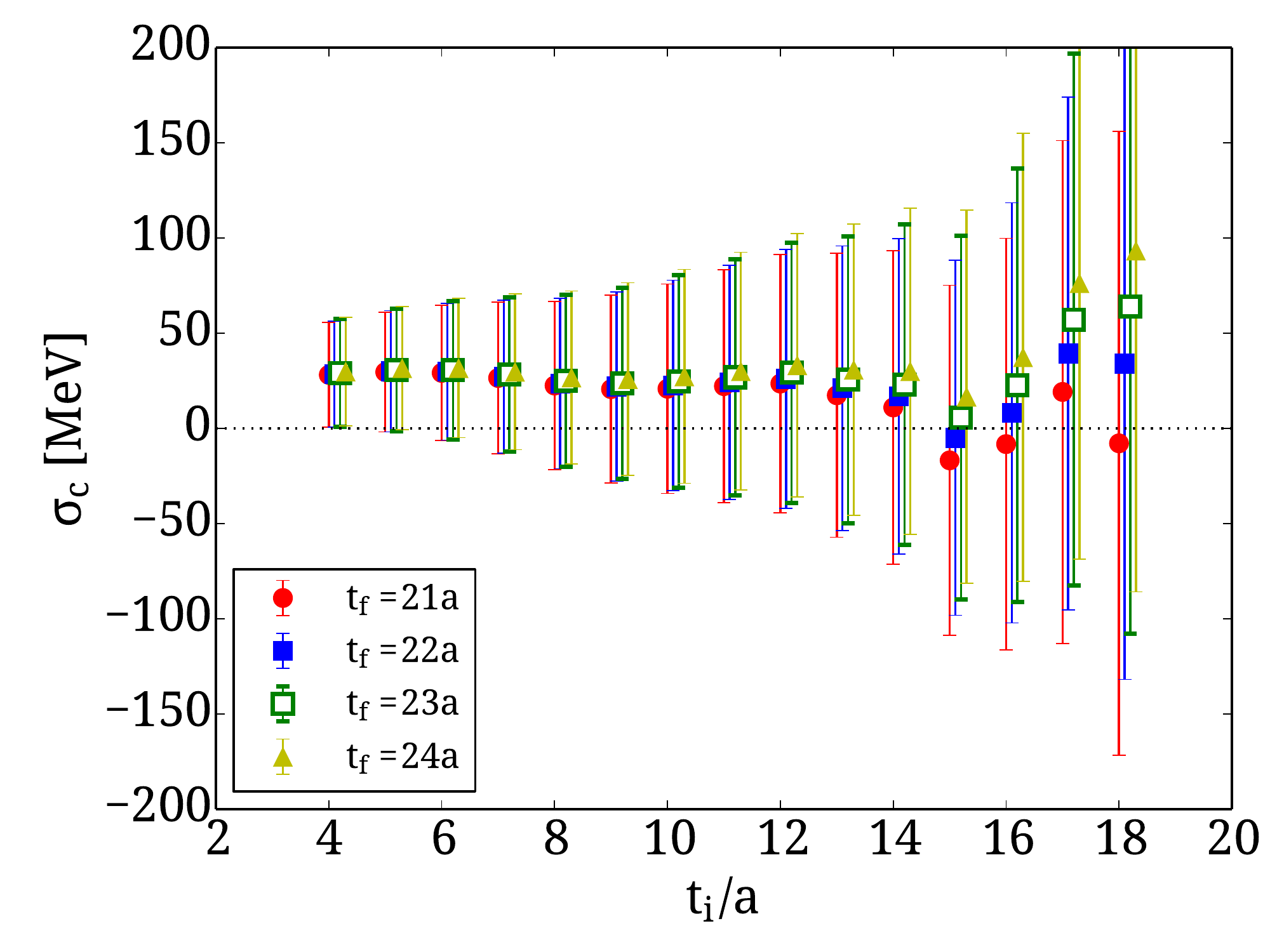} 
  \caption{The ratio from which the charm quark content of the nucleon, 
    $\sigma_{c}$, is extracted. The notation is the same as that of
    Fig.~\ref{sigmaLight}.\label{sigmaCharm}}
\end{figure}

\begin{figure}[h!]
  \includegraphics[width=0.8\linewidth,angle=0]{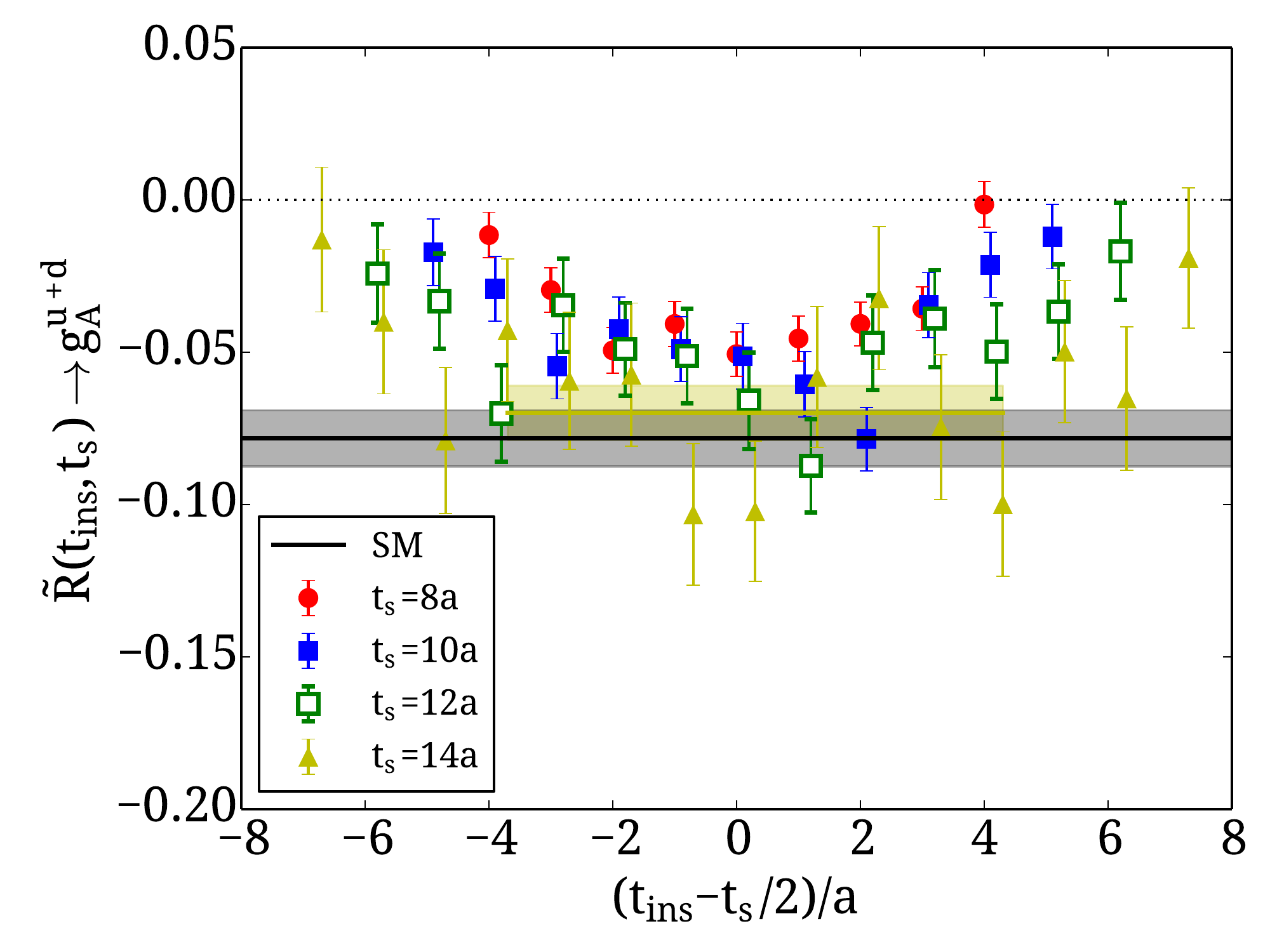} \\
  \includegraphics[width=0.8\linewidth,angle=0]{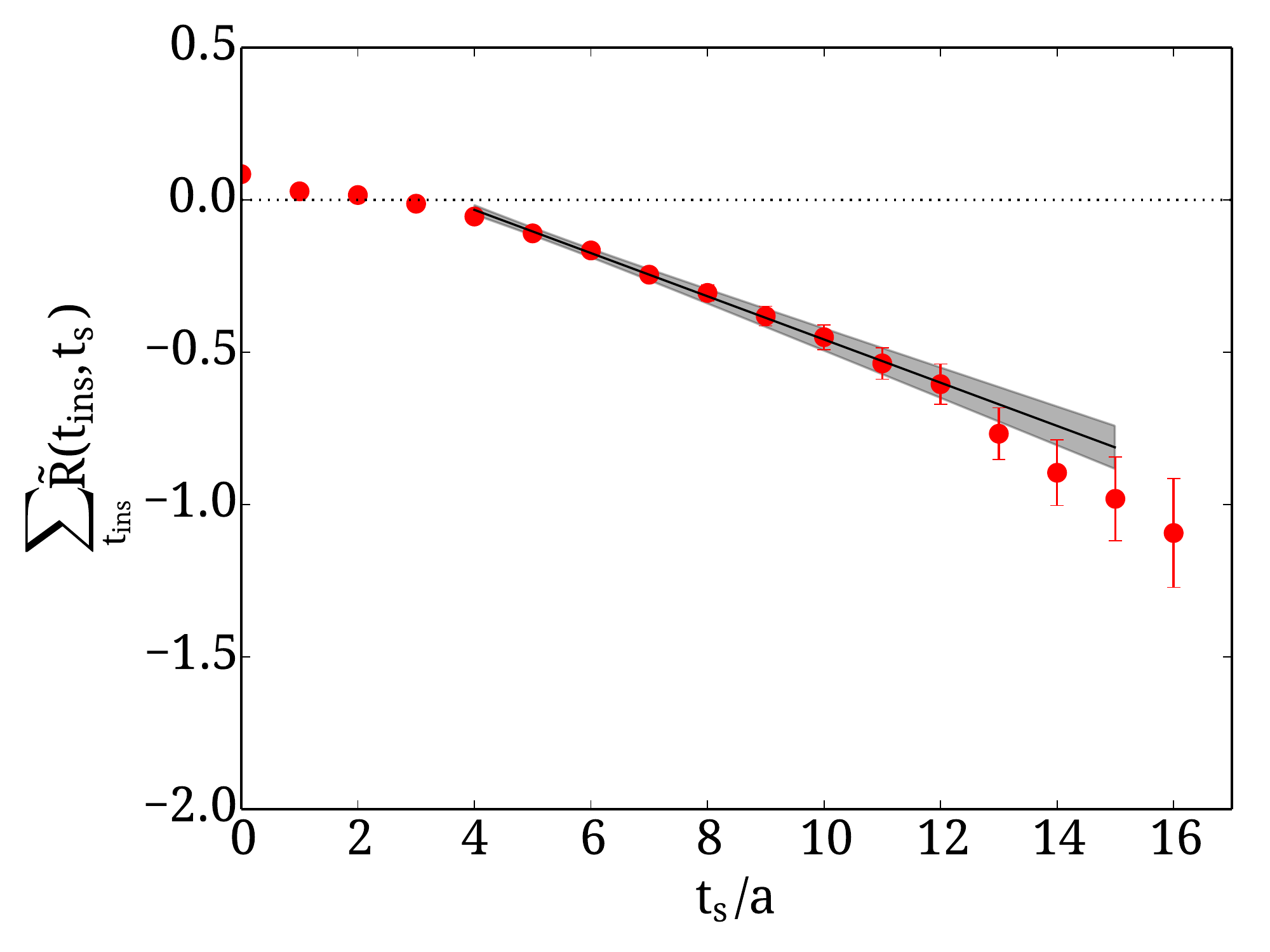} \\
  \includegraphics[width=0.8\linewidth,angle=0]{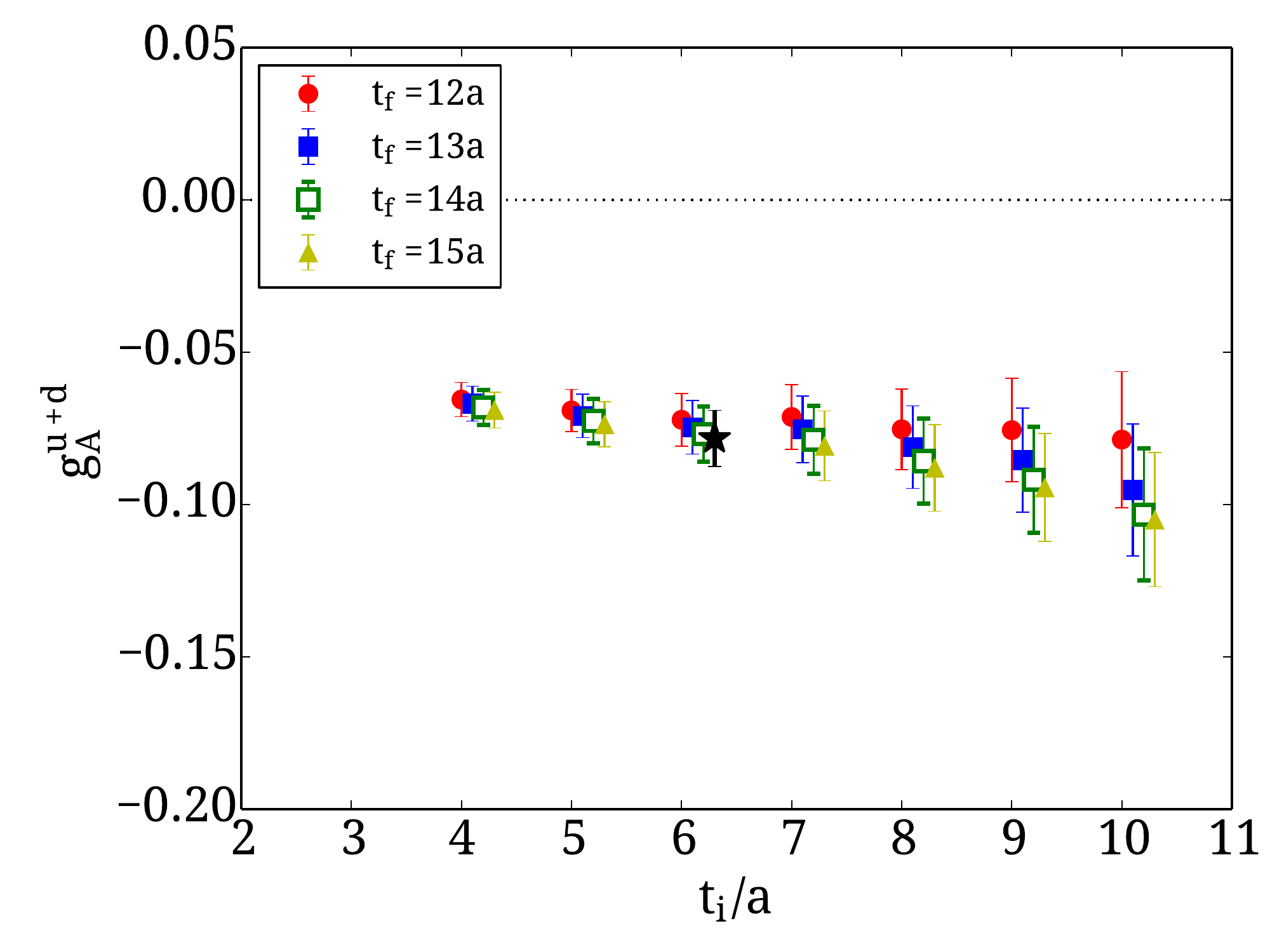} 
  \caption{The disconnected contribution to the renormalized ratio which yields the
    isoscalar axial charge of the nucleon, $g_A^{u+d}$. The upper
    panel shows the ratio as a function of the insertion time slice
    with respect to the mid-time separation ($t_{\rm ins}-t_s/2$) for
    source-sink separations $t_{\rm s}=8a$ (red filled circles), $t_{\rm
      s}=10a$ (blue filled squares), $t_{\rm s}=12a$ (green open squares) and
    $t_{\rm s}=14a$ (yellow filled triangles). The central panel shows the
    summed ratio and the lower panel the results obtained for the
    fitted slope of the summation method for various choices of the
    initial and final fit time slices as explained in the text. The
    star shows the choice of $t_i$, which yields the gray bands
    shown in the upper and central plots. \label{gALight}}
\end{figure}

\begin{figure}[h!]
  \includegraphics[width=0.8\linewidth,angle=0]{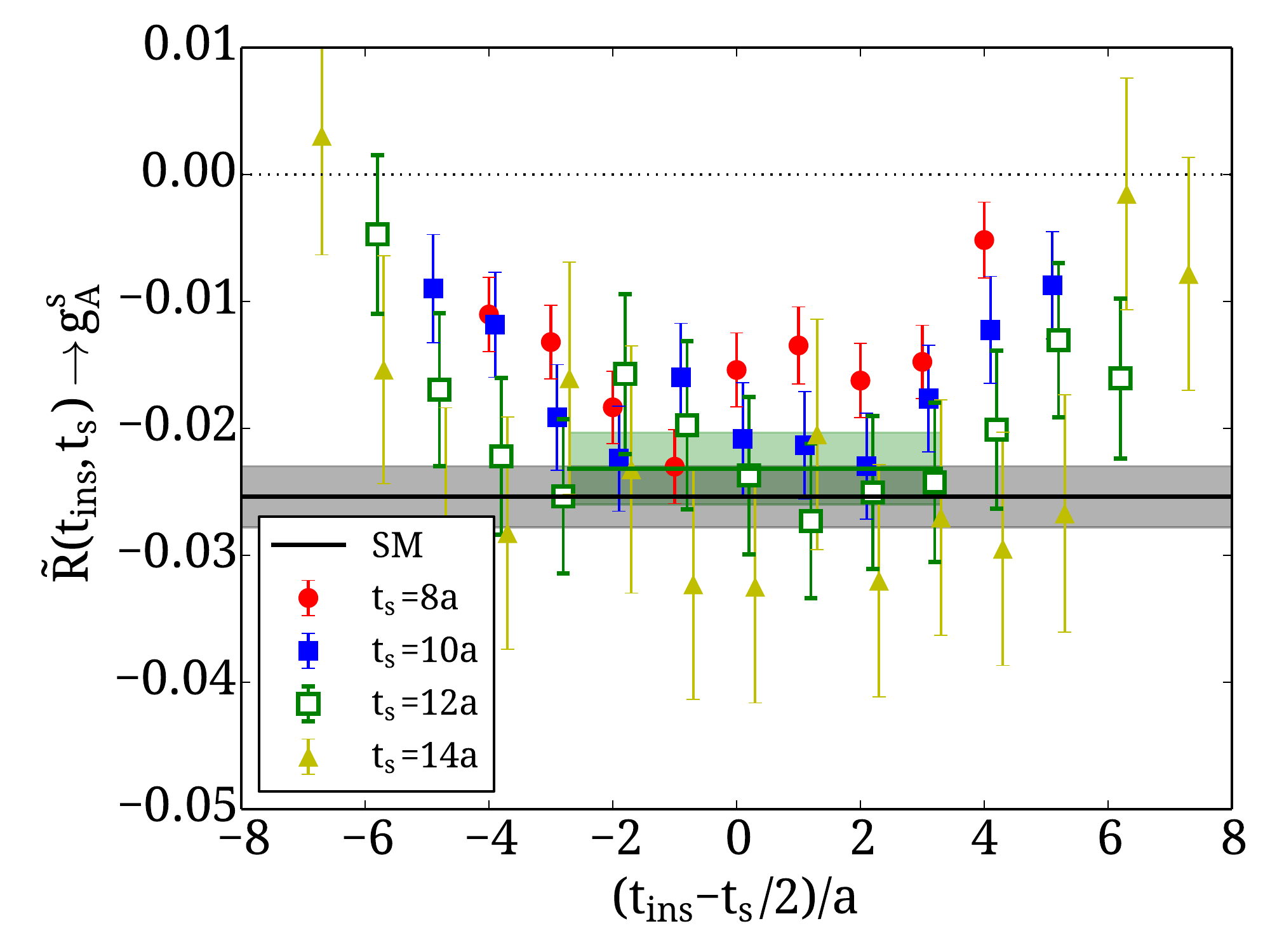} \\
  \includegraphics[width=0.8\linewidth,angle=0]{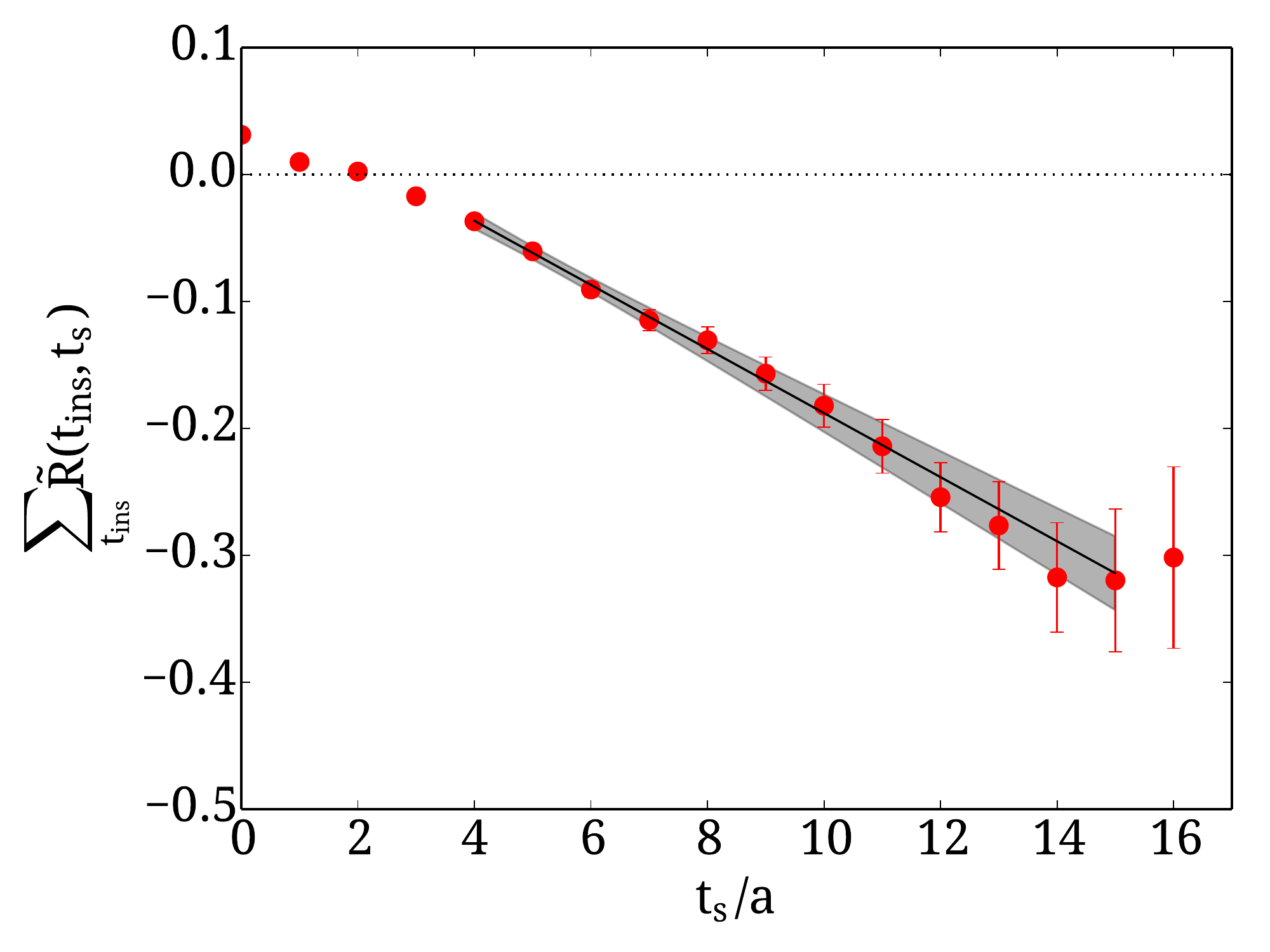} \\
  \includegraphics[width=0.8\linewidth,angle=0]{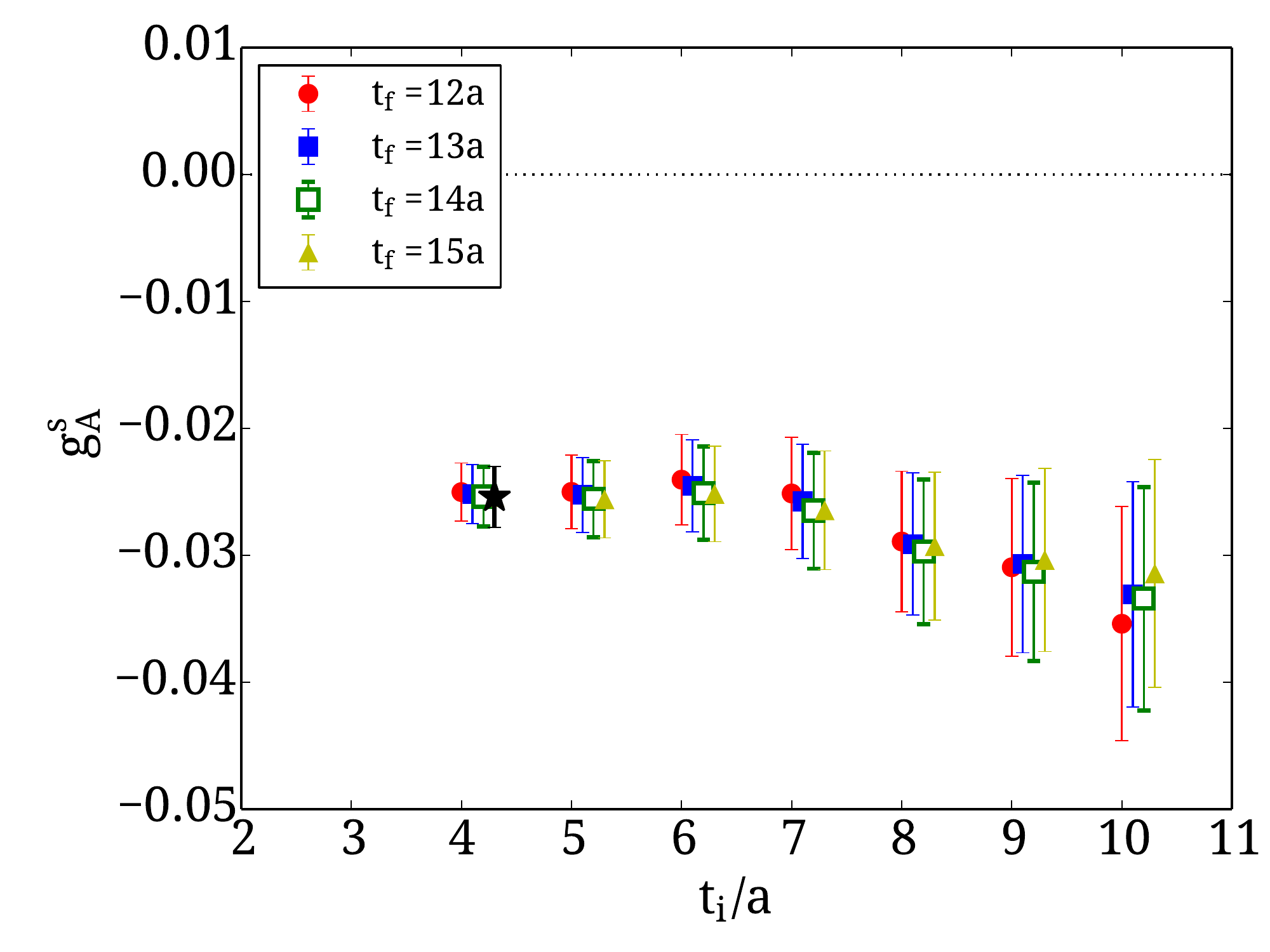} 
  \caption{The strange-quark contribution to the renormalized ratio yielding the nucleon axial
    charge $g_A^s$. The notation is the same as that of
    Fig.~\ref{gALight}.\label{gAStrange}}
\end{figure}

\begin{figure}[h!]
  \includegraphics[width=0.8\linewidth,angle=0]{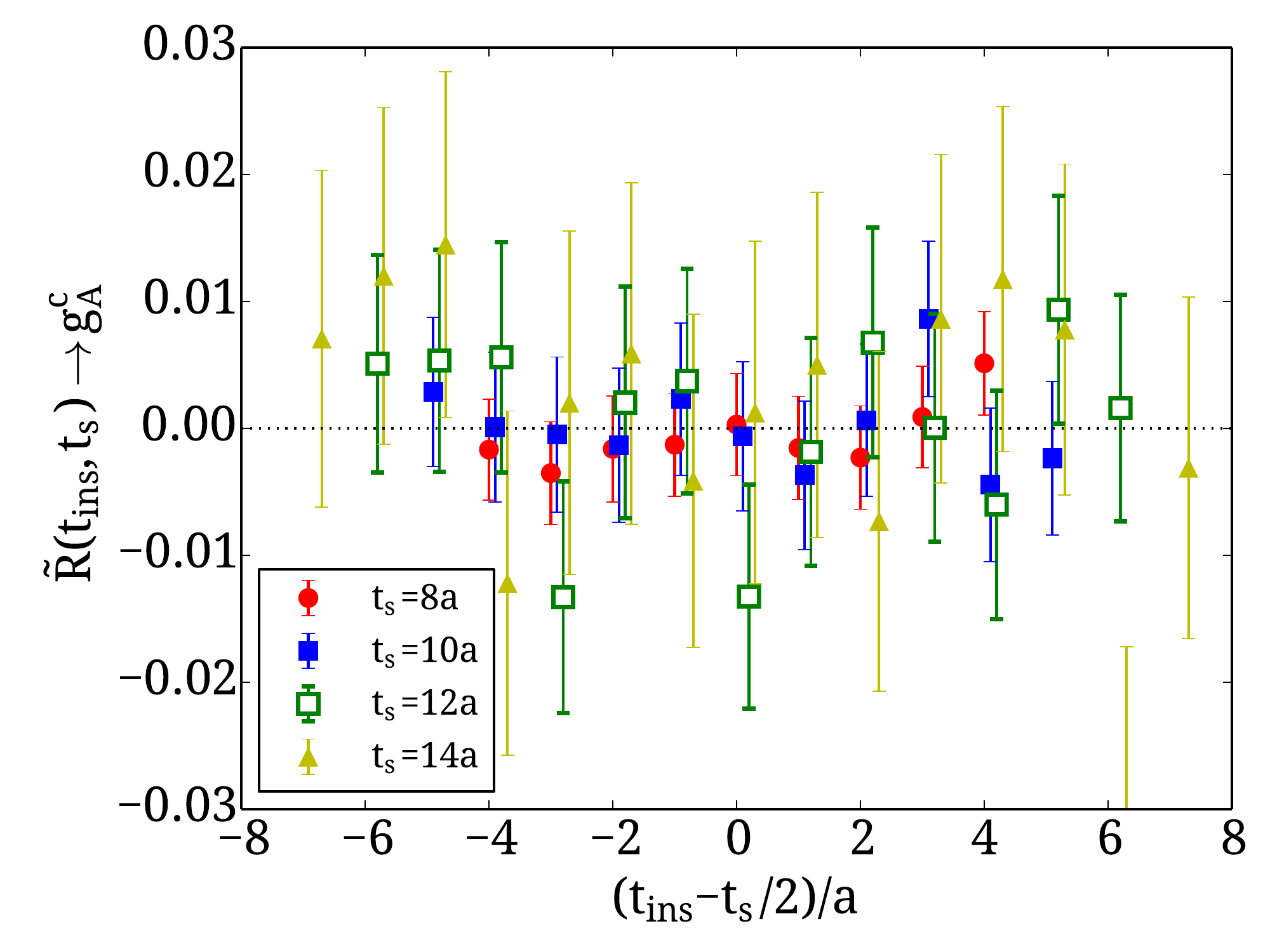} \\
  \includegraphics[width=0.8\linewidth,angle=0]{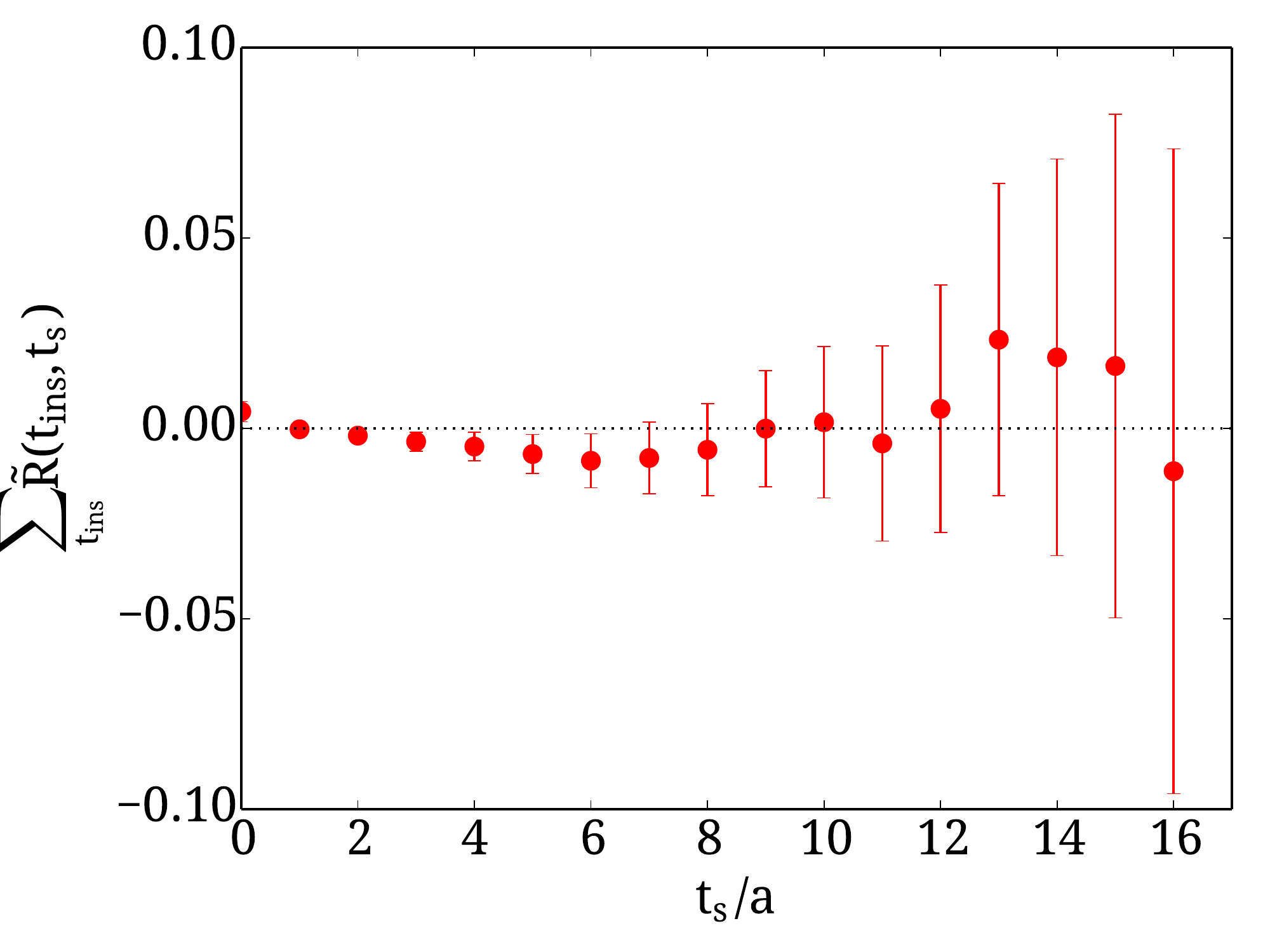} \\
  \includegraphics[width=0.8\linewidth,angle=0]{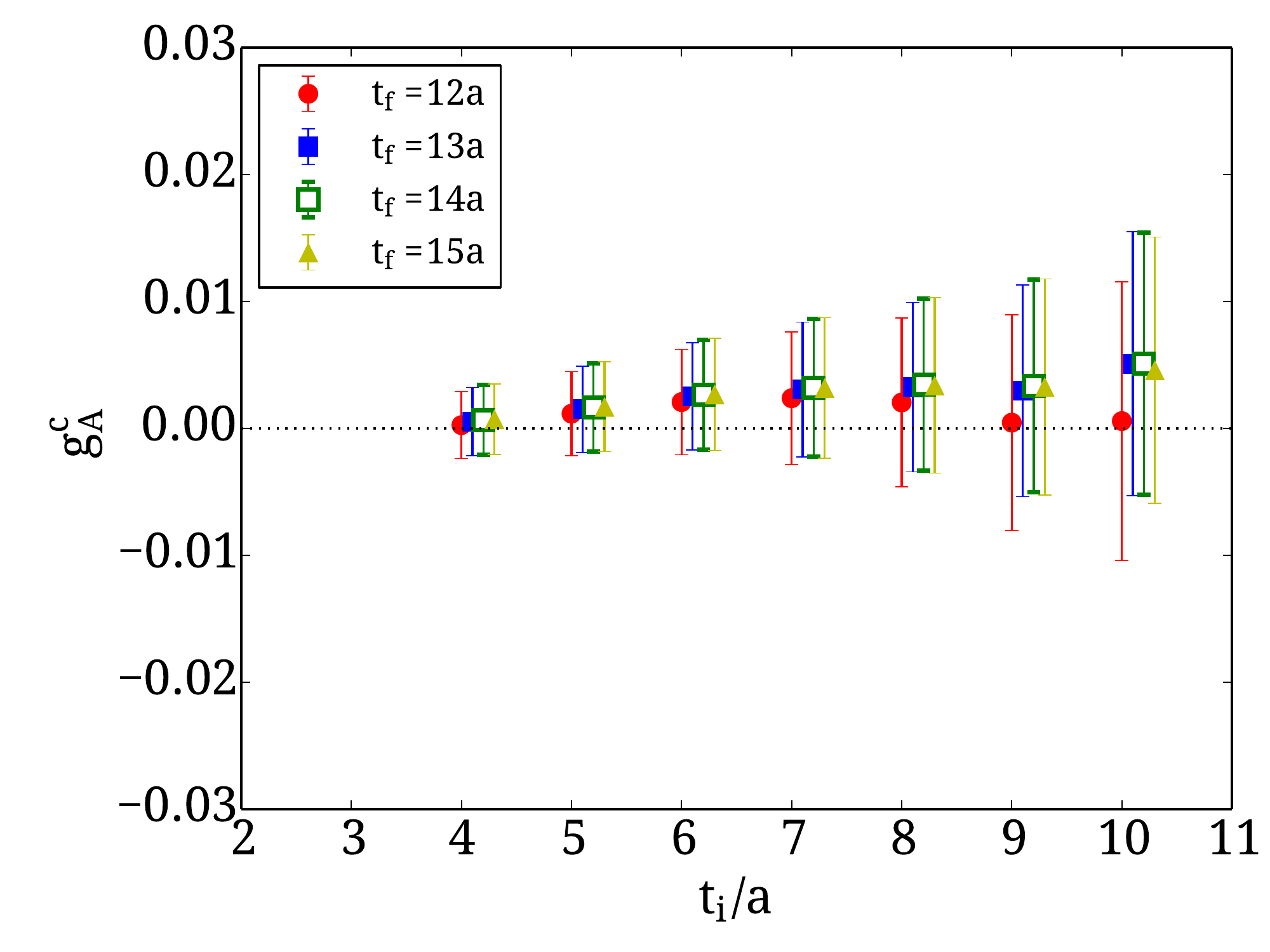} 
  \caption{The charm-quark contribution to the renormalized ratio yielding the nucleon
    axial charge $g_A^c$. The notation is the same as that of
    Fig.~\ref{gALight}.\label{gACharm}}
\end{figure}

\begin{figure}[h!]
  \includegraphics[width=0.8\linewidth,angle=0]{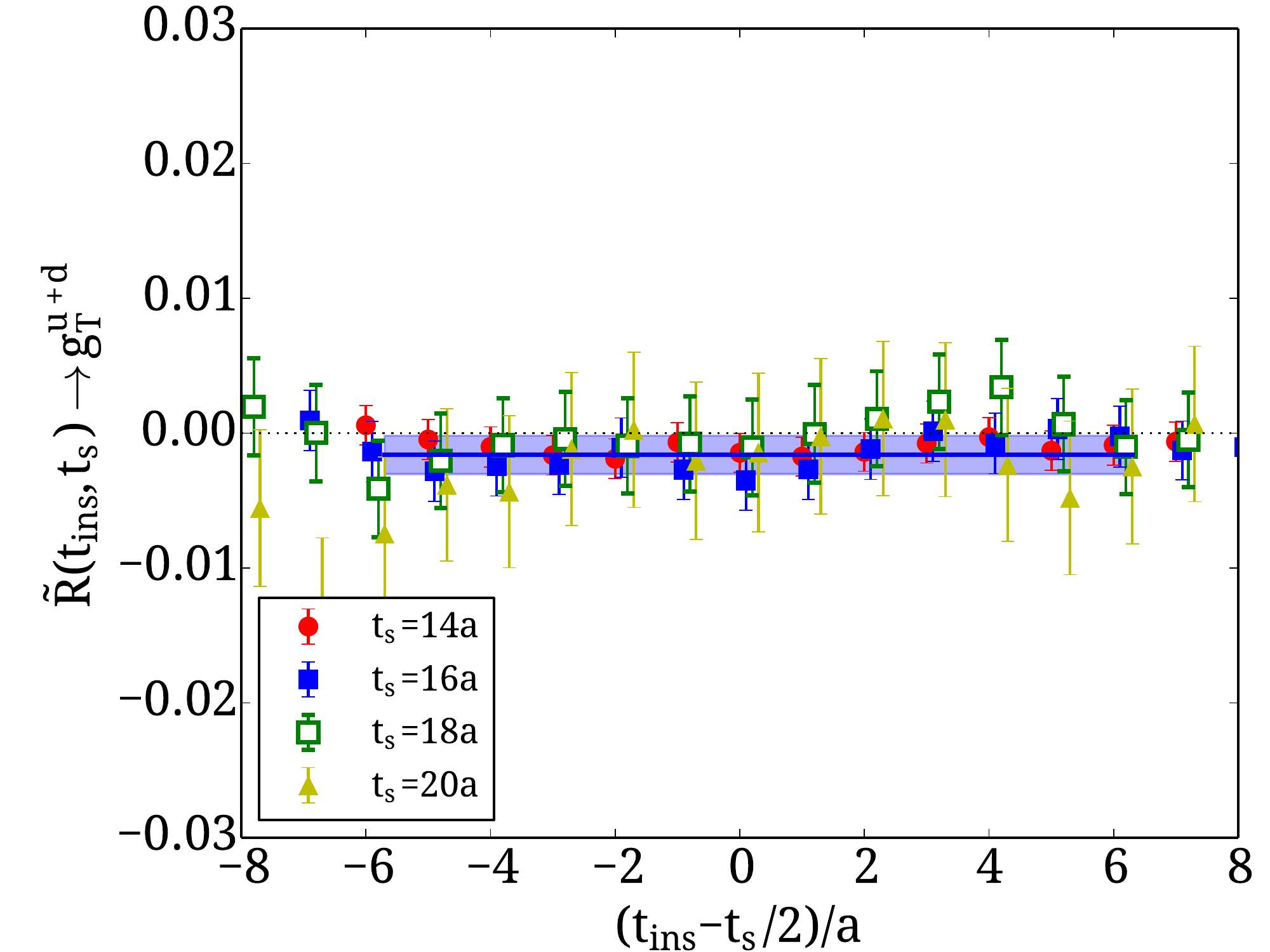}  \\
  \includegraphics[width=0.8\linewidth,angle=0]{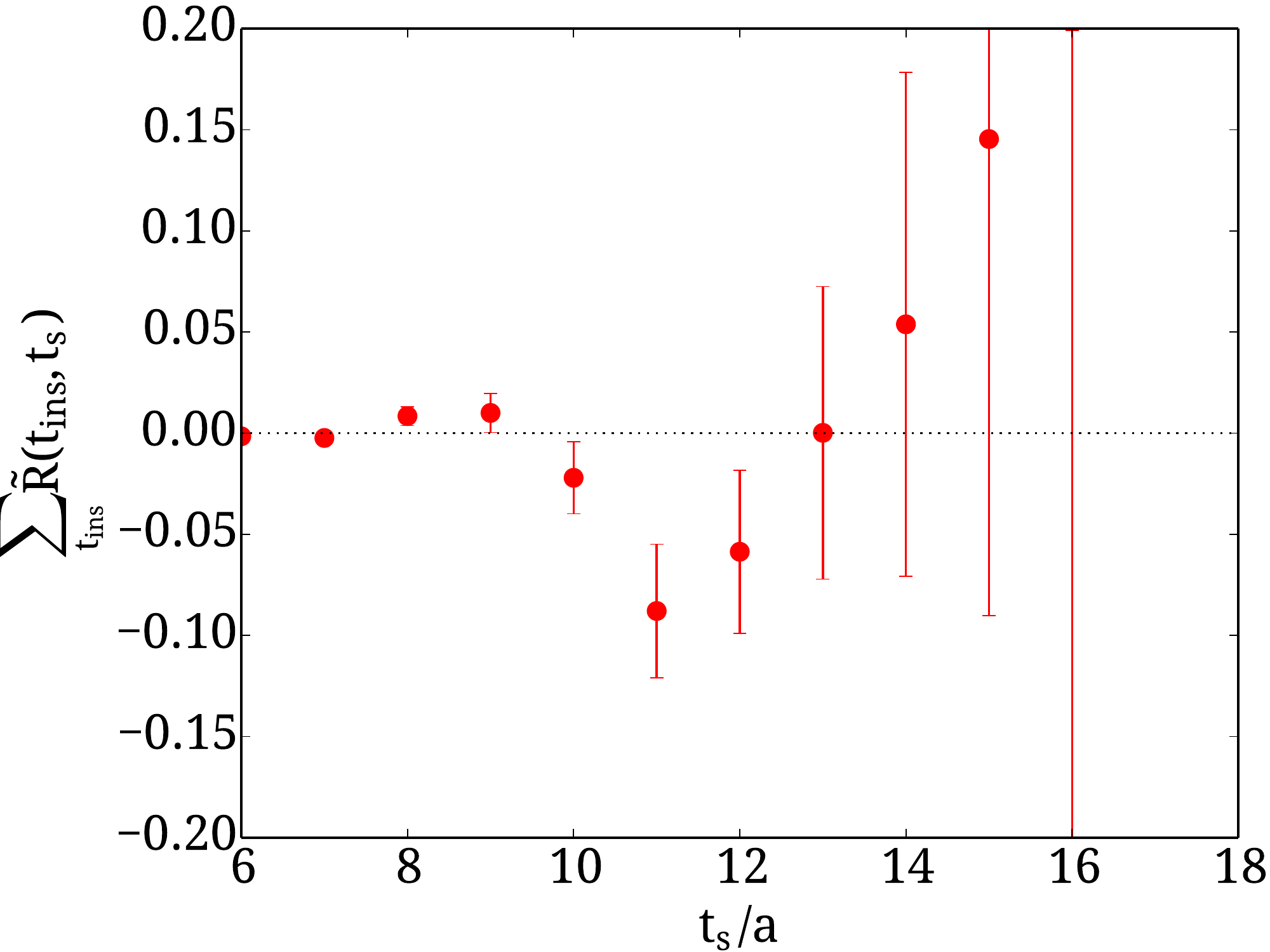} \\             
  \includegraphics[width=0.8\linewidth,angle=0]{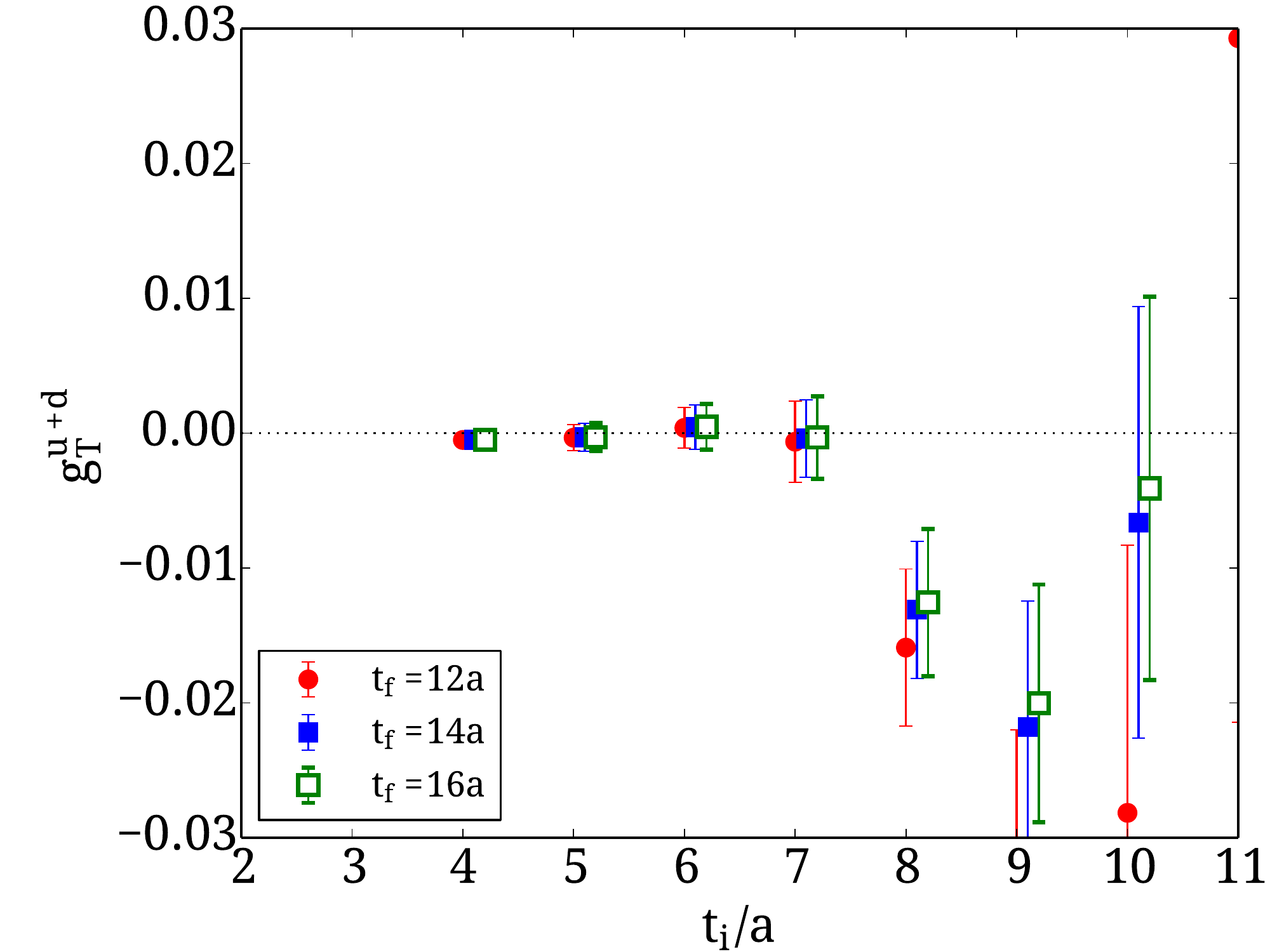}   
  \caption{The disconnected contribution to the renormalized ratio yielding the
    nucleon isoscalar tensor charge $g_T^{u +d}$. The notation is the
    same as that of Fig.~\ref{sigmaLight}.\label{A10T}}
\end{figure}

\begin{figure}[h!]
  \includegraphics[width=0.8\linewidth,angle=0]{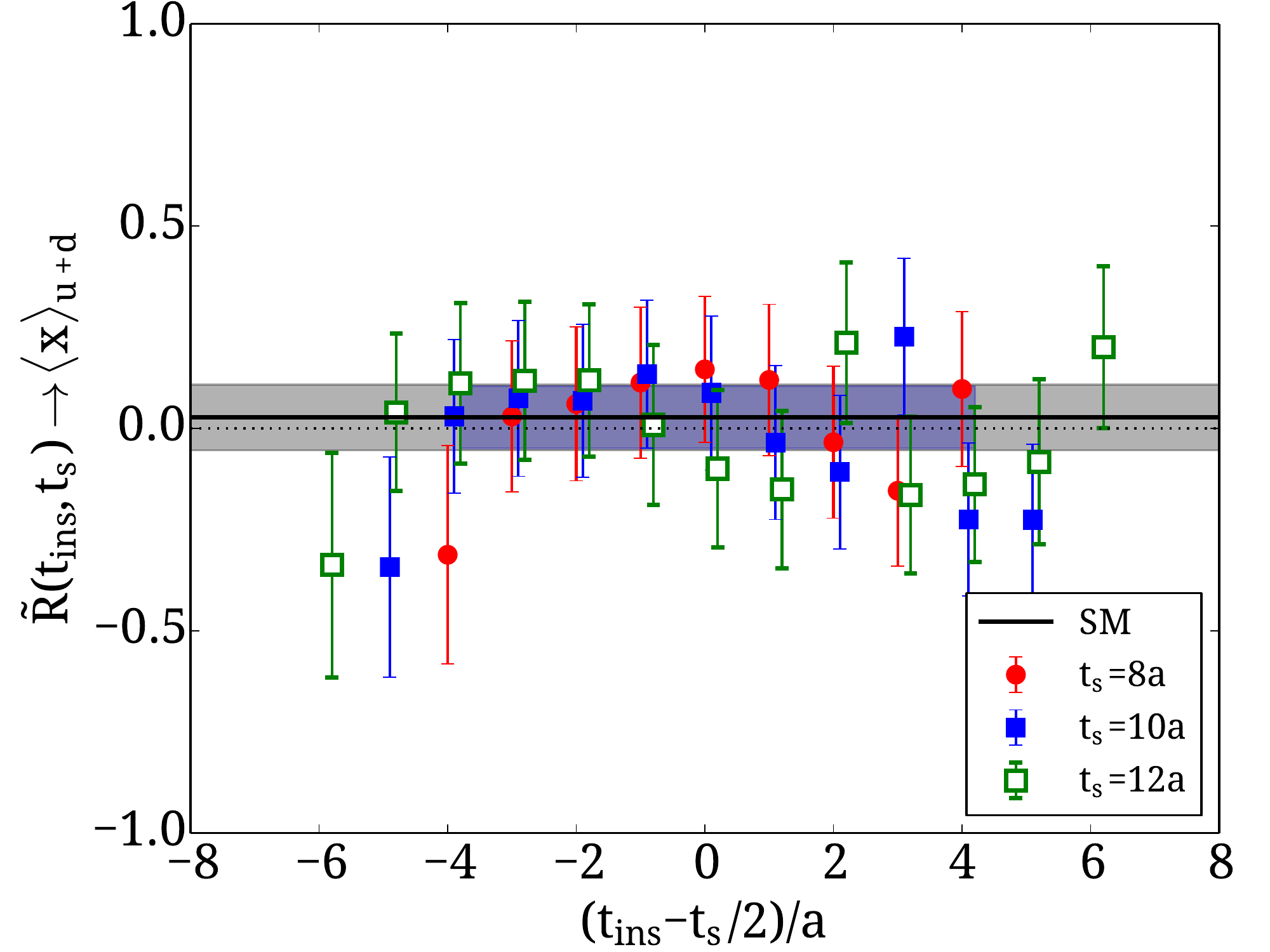}\\ 
  \includegraphics[width=0.8\linewidth,angle=0]{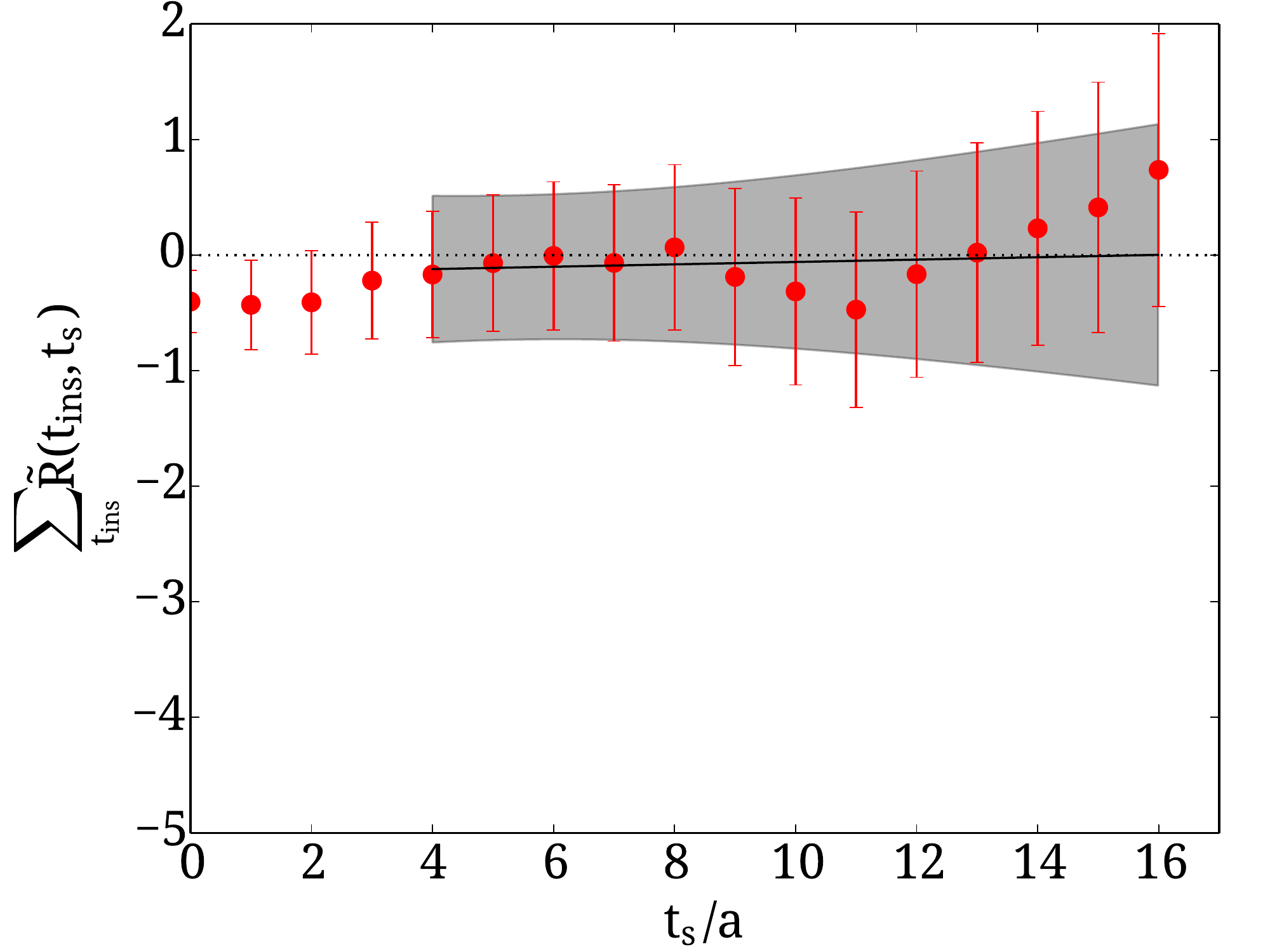}\\ 
  \includegraphics[width=0.8\linewidth,angle=0]{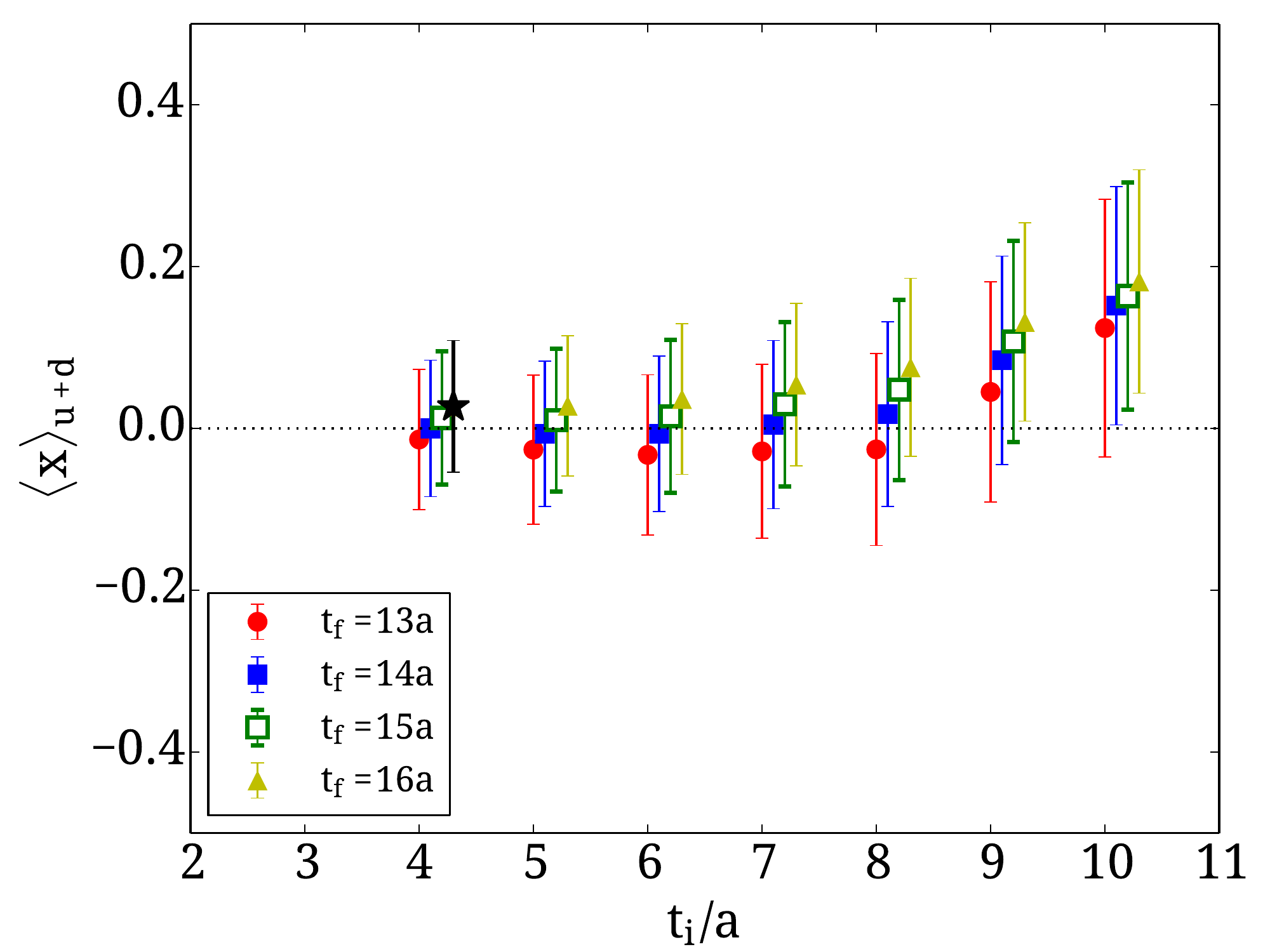} 
  \caption{The disconnected contribution to the renormalized ratio yielding the
    nucleon isoscalar momentum fraction $\langle x \rangle_{u+d}$. The
    notation is the same as that of Fig.~\ref{gALight}.\label{A20}}
\end{figure}

\begin{figure}[h!]
  \includegraphics[width=0.8\linewidth,angle=0]{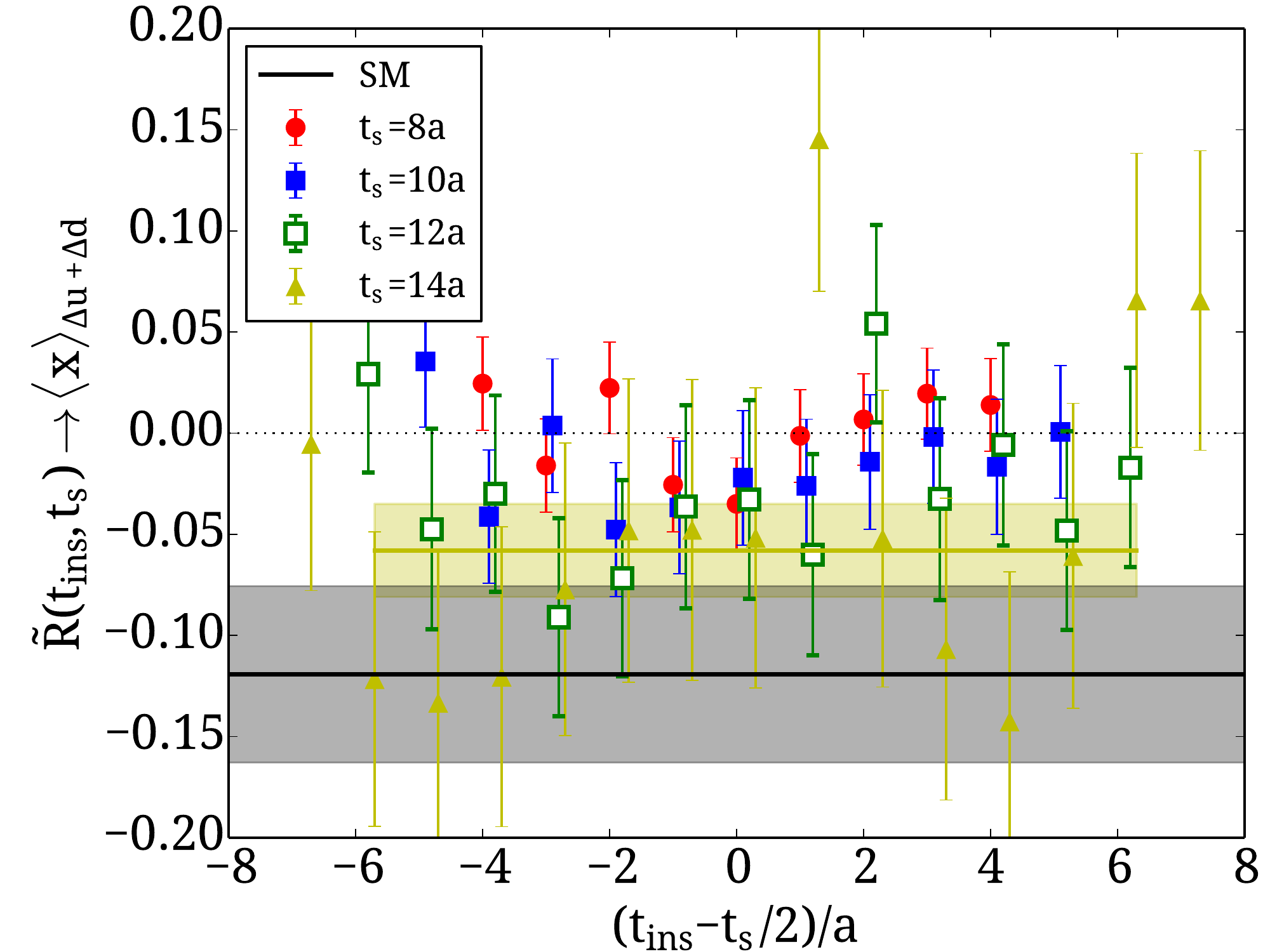}  \\
  \includegraphics[width=0.8\linewidth,angle=0]{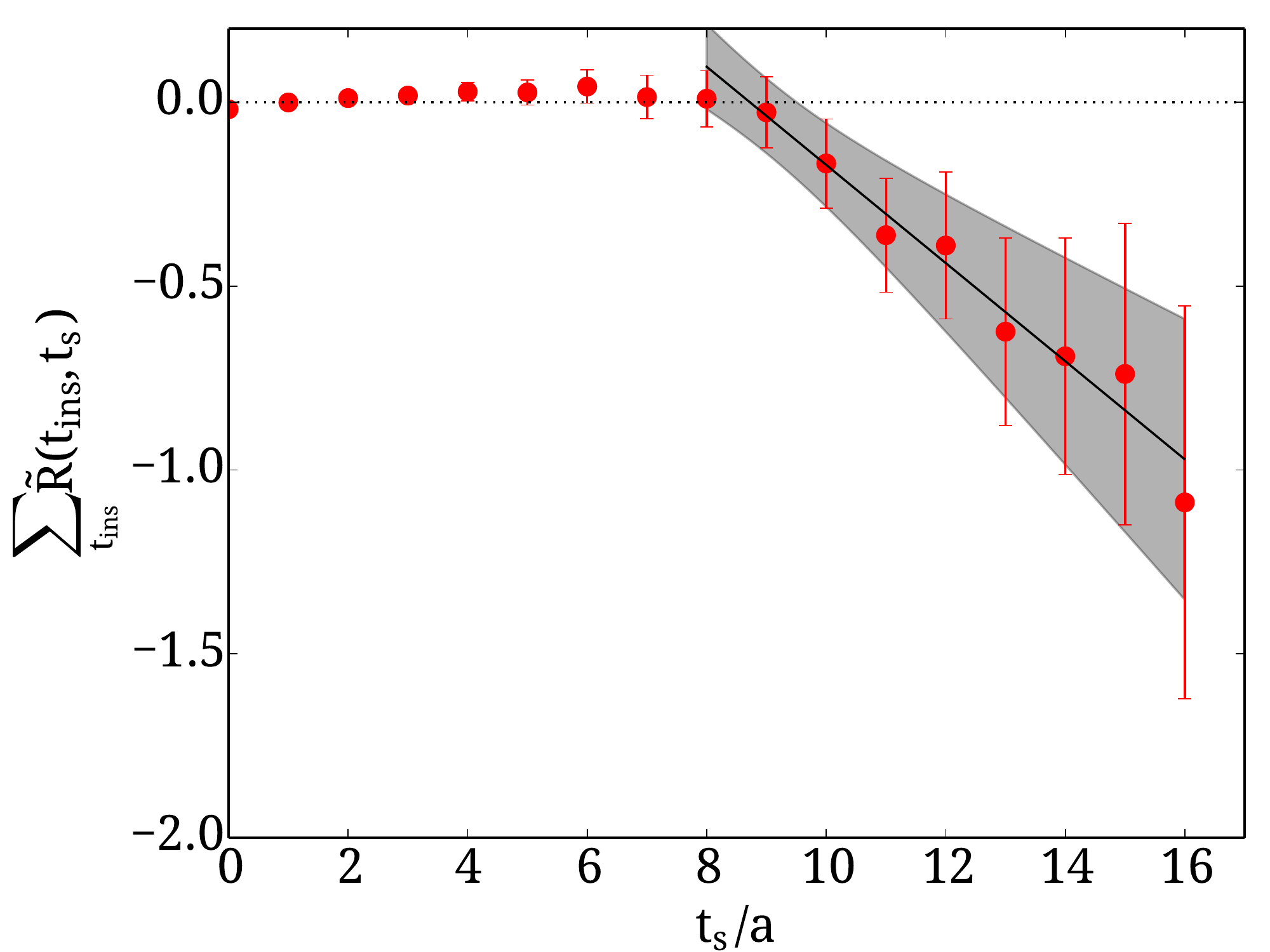} \\             
  \includegraphics[width=0.8\linewidth,angle=0]{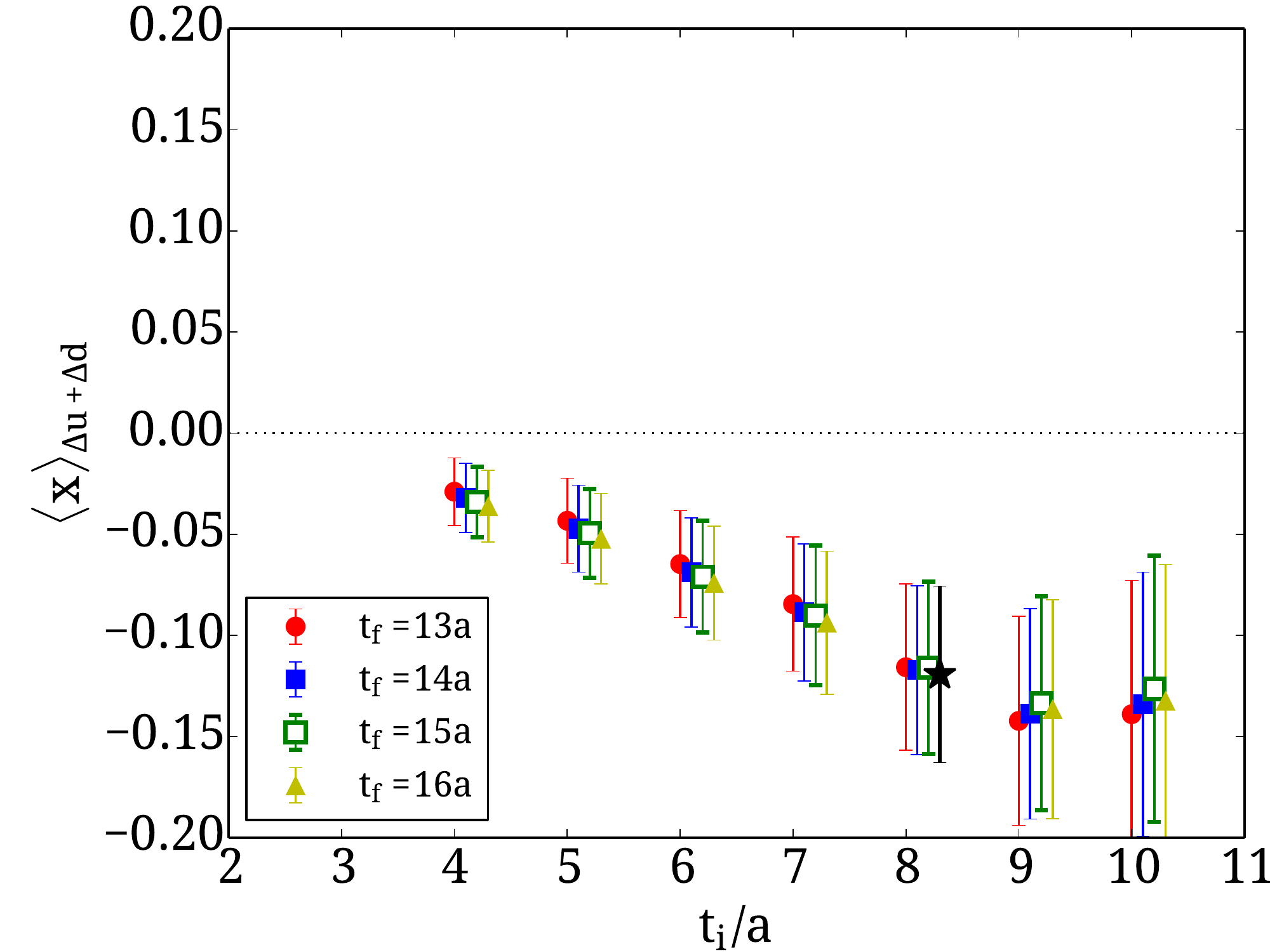}   
  \caption{The disconnected contribution to the renormalized ratio yielding nucleon isoscalar
    helicity moment $\langle x \rangle_{\Delta u + \Delta d}$. The
    notation is the same as that of Fig.~\ref{gALight}.\label{A20t}}
\end{figure}

\begin{figure}[h!]
  \includegraphics[width=0.8\linewidth,angle=0]{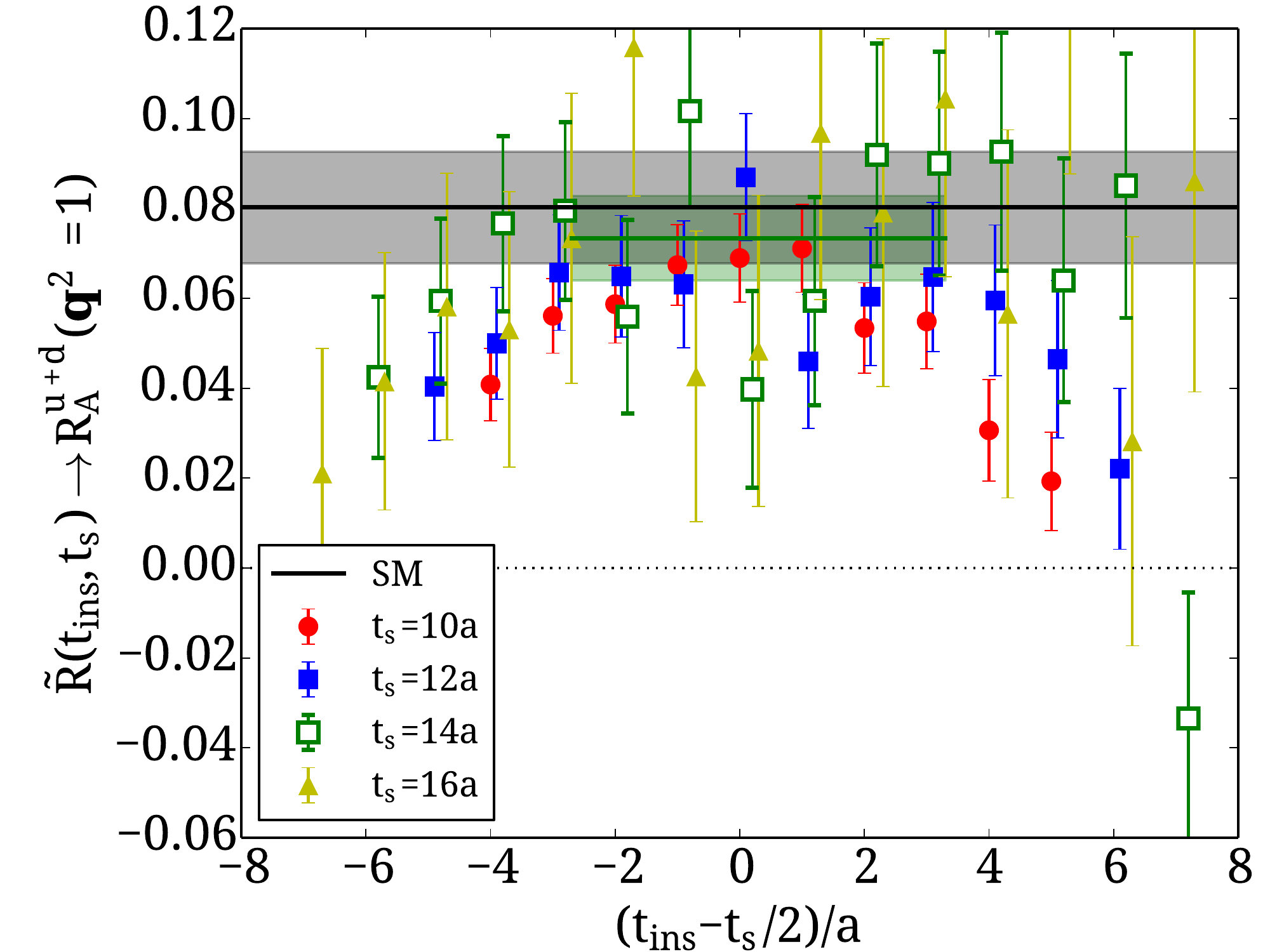}\\
  \includegraphics[width=0.8\linewidth,angle=0]{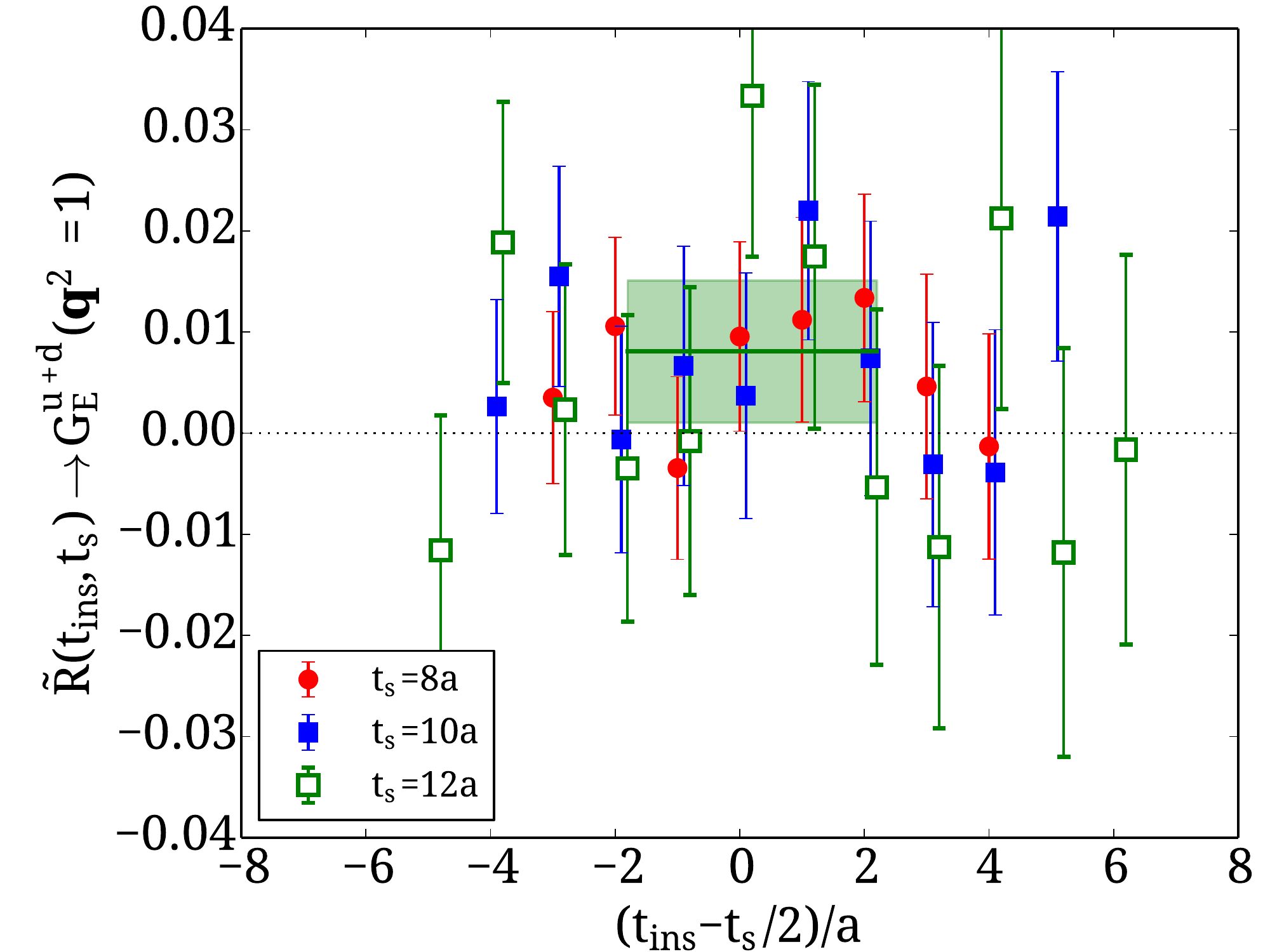}\\
  \includegraphics[width=0.8\linewidth,angle=0]{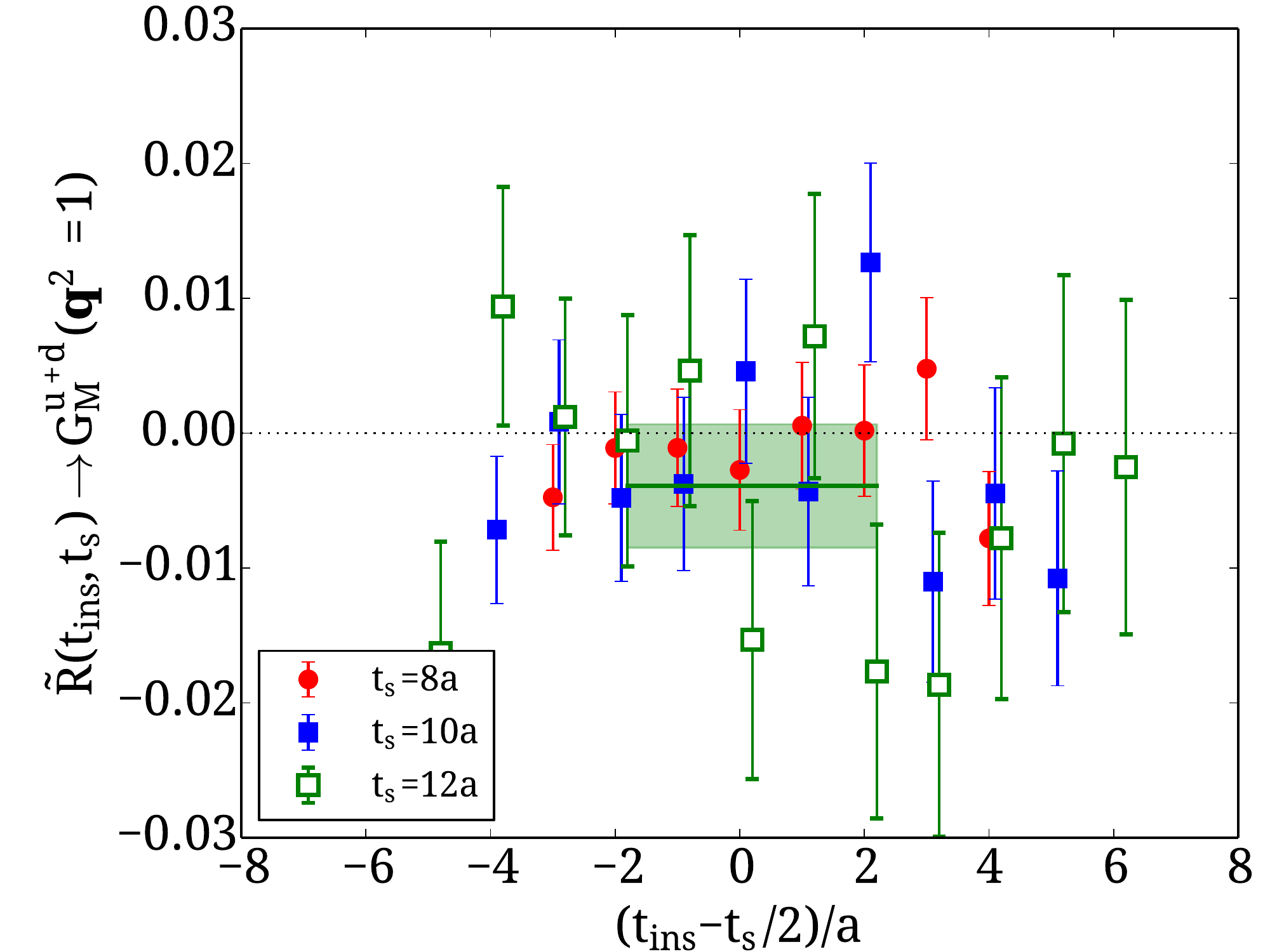}
  \caption{Disconnected contributions to the renormalized  ratio
yielding the isoscalar axial-vector and
    pseudo-scalar form-factors $G_A$ and $G_p$ (upper),  the
    electric form-factor $G_E$ (center) and  the magnetic
    form-factor $G_M$ (lower) at the
    lowest non-zero momentum transfer allowed for this lattice
    size. \label{finite_mom_plat}}
\end{figure}
\begin{figure}
  \includegraphics[width=1\linewidth,angle=0]{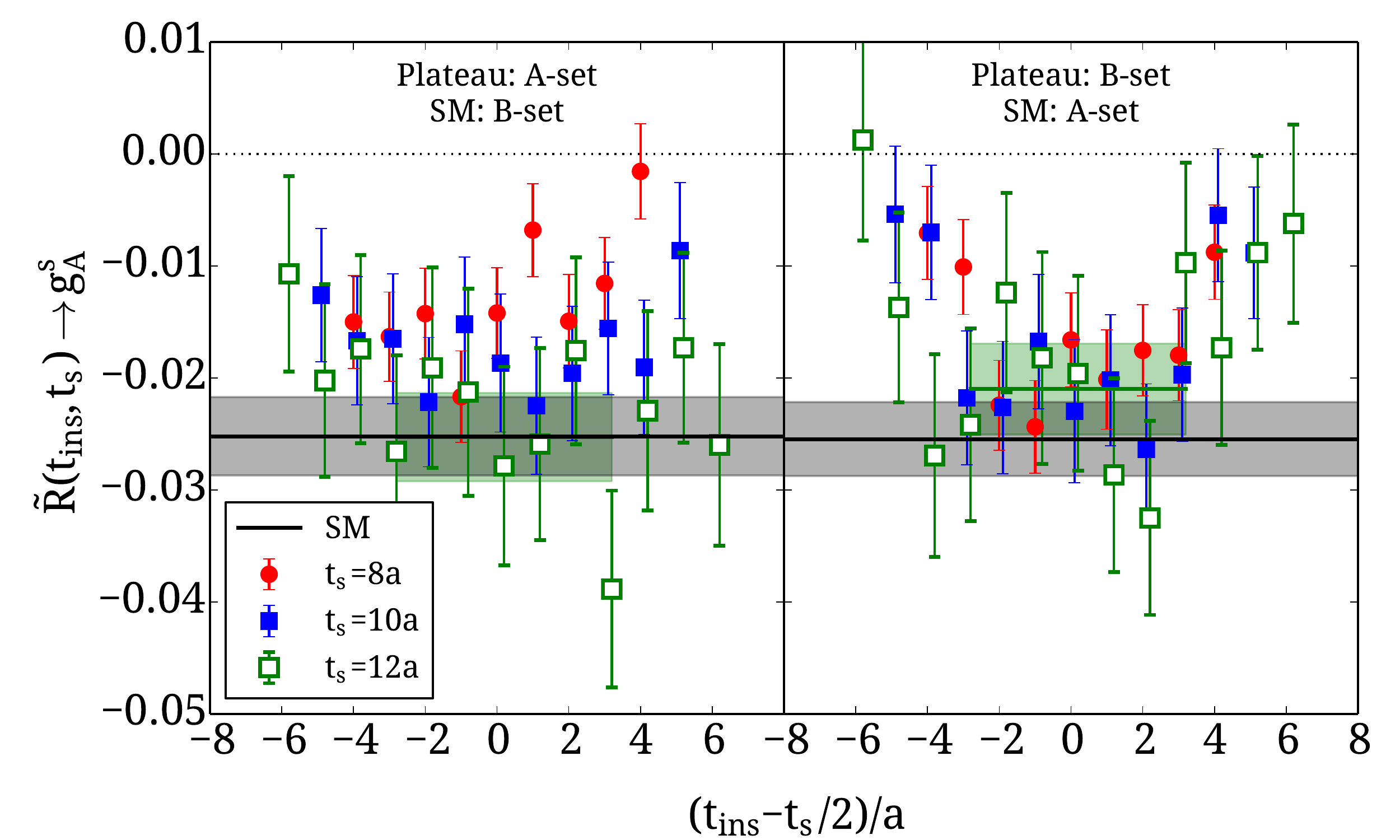}
  \caption{The renormalized ratio which yields the strange-quark
    contribution to the axial charge of the nucleon, $g_A^{s}$. In the
    left panel, the plateau method is used on the first half of
    the ensemble (A-set), while the summation method is used
    on the second half of the ensemble (B-set). In the right
    panel, he plateau method is used on the A-set, while the
    summation method is used on the B-set.}
  \label{gA_strange_parts}
\end{figure}
Before comparing the lattice matrix elements $R_{\rm GS}$ with experiment
we need to renormalize them. We  denote the renormalized ratio by $\tilde{R}(t_{\rm ins},t_s)$.
Regarding the renormalization of the
sigma terms, the twisted mass formulation has the additional
advantage of avoiding any mixing, even though we are using Wilson-type
fermions~\cite{Dinter:2012tt}.  For the case of the axial charge,
renormalization involves mixing from the three quark sectors. For the
tree-level Symanzik improved gauge action this mixing was shown to be
a small effect of a few percent~\cite{Skouroupathis:2008mf}. We expect
this to hold also for the Iwasaki action used in this work and for the other isoscalar quantities. In this work,  we neglect the small difference
in the renormalization constant between connected and disconnected contributions 
 and  we use
the same renormalization
constants as for the connected piece. They are given in Table~\ref{tab:renormalization constants}.  The value of $Z_P$ needs a pole subtraction and is taken from Ref.~\cite{ETM:2011aa, ETMC}, while all the others have been calculated using the approach given in Refs.~\cite{Alexandrou:2010me, Alexandrou:2012mt}. All the renormalization constants, except $Z_A$ which is scheme and scale independent, are
converted from RI-MOM to $\overline{\rm MS}$ at a scale of $\mu=2$~GeV. The conversion factors  for $Z_T$ are taken from Ref.~\cite{Gracey:2003yr},
and for the one-derivative operators from Ref.~\cite{Alexandrou:2010me}, computed to three-loops. We remark that in the twisted basis the scalar charge is renormalized with $Z_P$.


In Fig.~\ref{sigmaLight} we show the results for the disconnected
contribution to the ratio from which the $\sigma_{\pi N}$-term is
extracted. The ratio is plotted versus the time separation of the
current insertion $t_{\rm ins}$ from the source, shifted
by $t_s/2$. When this ratio becomes time independent (plateau region)
fitting to a constant yields $\sigma_{\pi N}$. As can be seen, however,
increasing the source-sink time separation increases the value extracted
from fitting to the plateau (plateau value).  We observe that one
requires a source-sink time separation of at least 18 to 20 time slices in
order for the plateau value to stabilize. This is a distance of
$\gtrsim 1.5$~fm, which is significantly larger than the nominal
source-sink separation of 1.0~fm-1.2~fm typically used in nucleon
matrix element calculations. In the central panel we show the ratio summed over
the insertion time slice as given in Eq.~\eqref{RatioSumDet} referred
to as summation method (SM) as a function of the source-sink time
separation time. As explained earlier, by fitting the ratio to a
straight line one obtains the desired matrix element as the
slope. This is done for several choices of the initial and final fit
time slices ($t_i$ and $t_f$ respectively) with the results displayed
in the lower panel of the figure.  As one increases the initial fit
time slice the excited state contributions are expected to become
smaller and thus the fitted value stabilizes. Note, however, that the
slope changes and one needs to vary the fit range until the slope
converges.  Therefore, if one has only a small number of source-sink time
separations one may miss the variation of the slope. As in the case of
the plateau method where we take the smallest $t_s$ for which excited
states are sufficiently suppressed, it is desirable to take the
smallest $t_i$ for which the excited states no longer contribute
significantly, since the error to signal ratio increases with
$t_i$. Taking the value of the slope to be the one given by the star yields the value of $\sigma_{\pi N}$ shown by the gray band in
the upper panel of the figure.  As can be seen, the resulting value is
in agreement with the (colored) band obtained from the plateau method
for $t_s/a=20$.

A similar analysis is undertaken for the strange- and charm-quark
sigma terms, shown in Figs.~\ref{sigmaStrange} and~\ref{sigmaCharm}
respectively. For $\sigma_s$, similar remarks can be made as in the
case of $\sigma_{\pi N}$, most notably concerning the large
source-sink separation required for the plateau method to converge. As
expected, the results between the summation and the plateau method are
consistent also in this case, when excited states are suppressed.
Non-zero results for $\sigma_s$ were also obtained in
Ref.~\cite{Gong:2013vja} using optimal noise sources and low-mode
substitution techniques.  For the case of the charm content, our
results are consistent with zero both when using the plateau method as
well as when using the summation method allowing us only to obtain an
upper bound to its value. In Ref.~\cite{Gong:2013vja} a non-zero
result was obtained as one approaches the chiral limit. Since our aim
in this work is to compute quark loops using high statistics for one
ensemble we will address the quark mass dependence in a follow-up
work.

Similar analyses are carried out for the disconnected contributions
entering the ratios determining the nucleon axial charge.   For observables like $g_A$
where one does not have the $\tau^3$ flavor combination in the twisted basis 
it is advantageous to use the discrete symmetries of the
twisted mass formulation~\cite{Frezzotti:2003ni,Frezzotti:2004wz}, namely parity combined with isospin flip
$u\leftrightarrow d$, $\gamma_5$-isospin hermiticity, and
charge-$\gamma_5$-isospin hermiticity, in order to reduce
gauge noise. Considering the properties of the
quark loops and of the  nucleon two-point functions that enter in the
computation of the disconnected three-point function under these
symmetries one can derive appropriate products taking their real or
imaginary parts thus suppressing gauge noise. This was shown to be
advantageous in  the calculation of the first moments of
the unpolarized momentum distribution in Ref.~\cite{Deka:2008xr}.
These symmetries are used for the results shown from now on.  In
Figs.~\ref{gALight}, \ref{gAStrange} and~\ref{gACharm} we show,
respectively, results for the ratio from which the nucleon matrix 
elements of the axial-vector current yielding the isoscalar $g_A$,the strange $g_A^s$ and the charm $g_A^c$ are extracted.  We
first note that for the case of $g_A^{u+d}$ we observe less contamination
from excited states than in the case of the sigma terms. This is
evident from the smaller source-sink time separations required in order for
the plateau or summation method to converge. Furthermore, we clearly
observe a non-zero value for the case of the disconnected
contributions to the isoscalar $g_A$ as well as for $g_A^s$. For $g_A^c$ our results are consistent with zero and we can only give
an upper bound to its value. The nucleon tensor charge $g_T^{u+d}$ is also
computed and the ratio from which is extracted is shown in Fig~\ref{A10T}. We observe a very small value
for the disconnected contribution, with an error of about 90\%. For the
summation method the statistical uncertainty does not allow a
meaningful fit.

The nucleon matrix elements involving derivative operators probe
moments of parton distributions, which are extracted from deep inelastic 
scattering measurements.  In this work we compute the disconnected
contributions to the isoscalar nucleon momentum fraction $\langle x
\rangle_{u+d}$, which involves the vector derivative operator and the
isoscalar nucleon polarized moment $\langle x \rangle_{\Delta u+\Delta
  d}$ involving the axial-vector derivative operator. 
We apply the symmetries of the twisted mass action discussed above
as well as  consider a moving
frame and thus have the nucleon carrying non-zero equal initial and
final momentum for three-point functions with
zero momentum transfer. We find that, when the nucleon carries the lowest momentum allowed for this lattice, the statistical error is reduced.
 The disconnected contributions to the
ratios, from which the matrix elements of the vector and axial-vector derivative operators, are extracted are shown in
Figs.~\ref{A20} and~\ref{A20t} respectively. For $\langle x
\rangle_{u+d}$ we find a value consistent with zero both with the
plateau and summation method. Having one unit of momentum improves the
signal enabling us to deduce an upper bound on the value of this matrix element. For $\langle x \rangle_{\Delta u+\Delta d}$ the
statistical errors remain large but nevertheless we obtain a non-zero
value. Considering a moving nucleon leads in this particular case
 to a substantial
reduction in the error. We note that increasing the sink-source time separation is
crucial in order for this observable to develop a non-zero
result. This is clearly seen in the slope which becomes non-zero for
$t_s/a>8$. Since a large $t_s$ also leads to larger errors it is no
surprise that such a large number of statistics is needed to obtain a meaningful signal. This may also indicate that even larger number of statistics are needed
to stabilize further the signal.

Apart from matrix elements for zero momentum transfer presented so
far, disconnected contributions arise in the isoscalar electromagnetic
and axial form factors at finite momentum. Computationally, these are
straightforward to extract, since one takes the Fourier transform of
the insertion coordinate of the loop to obtain the matrix element at
all momenta.  The finite momentum matrix elements, however, are
expected to be nosier than the zero-momentum ones, since the energy
factors appearing in the exponents of the signal are larger. The
disconnected contributions to the axial form-factors, electric
form-factor and magnetic form-factor are shown in
Fig.~\ref{finite_mom_plat} for a single unit of momentum transfer. Due to the
structure of the matrix elements and the way these are computed on the
lattice, for the case of the axial form factors 
$G_A$ and $G_p$, the plot shows the ratio of a linear
combination from which these form factors are extracted after
the plateau fit. $G_E$ and $G_M$, on the other hand, can be extracted
from different ratios allowing us to plot them separately. We note
that we perform a similar analysis for these quantities as for the
zero-momentum case where both plateau and summation methods are
investigated for the optimal fit ranges. For the axial form-factors we
obtain a clearly non-zero value. For the electromagnetic case, the
disconnected contributions for both the isoscalar electric and
magnetic form factors are statistically consistent with zero.

\begin{figure}[h!]
  \includegraphics[width=0.8\linewidth,angle=0]{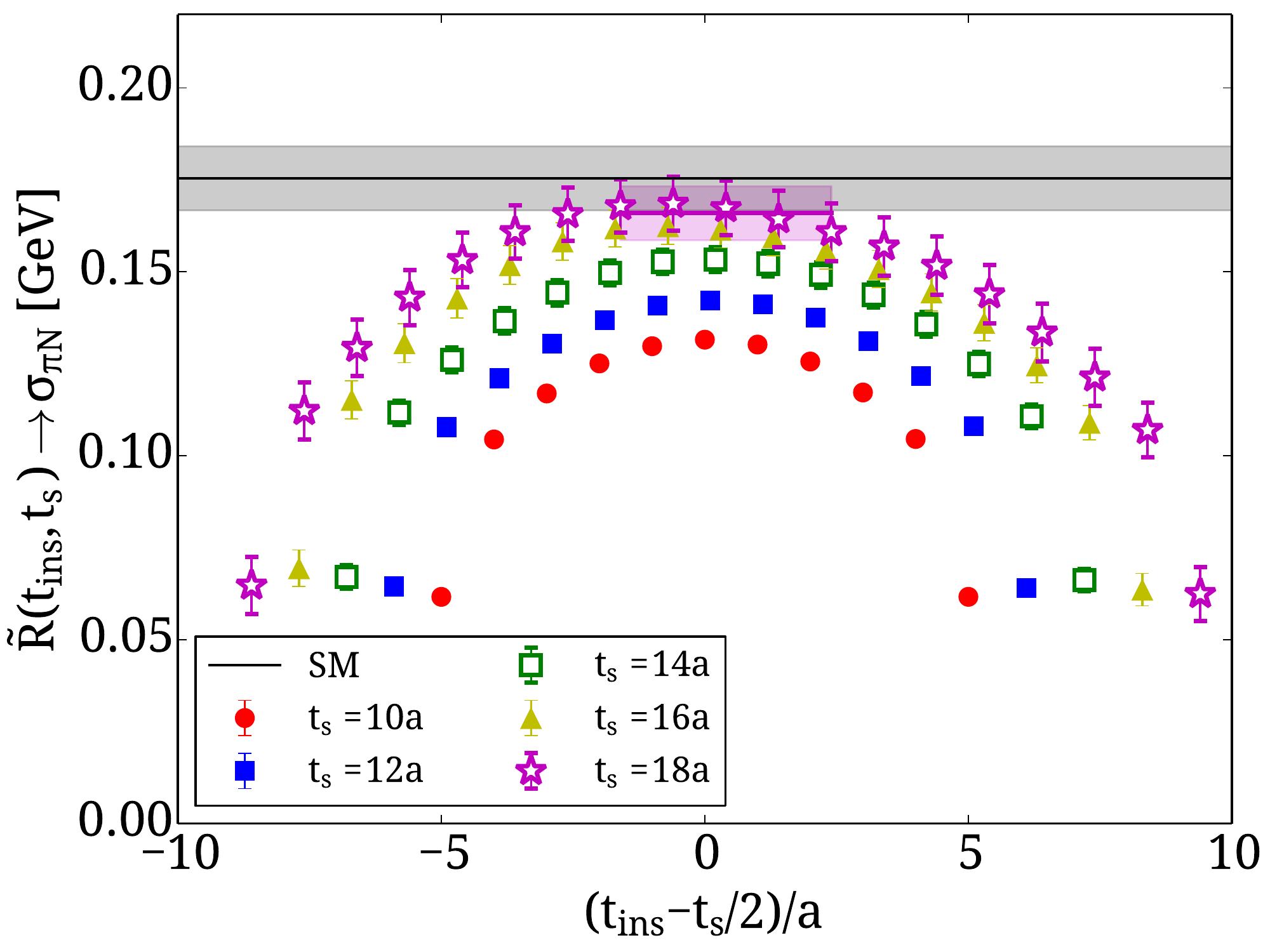}\\
  \includegraphics[width=0.8\linewidth,angle=0]{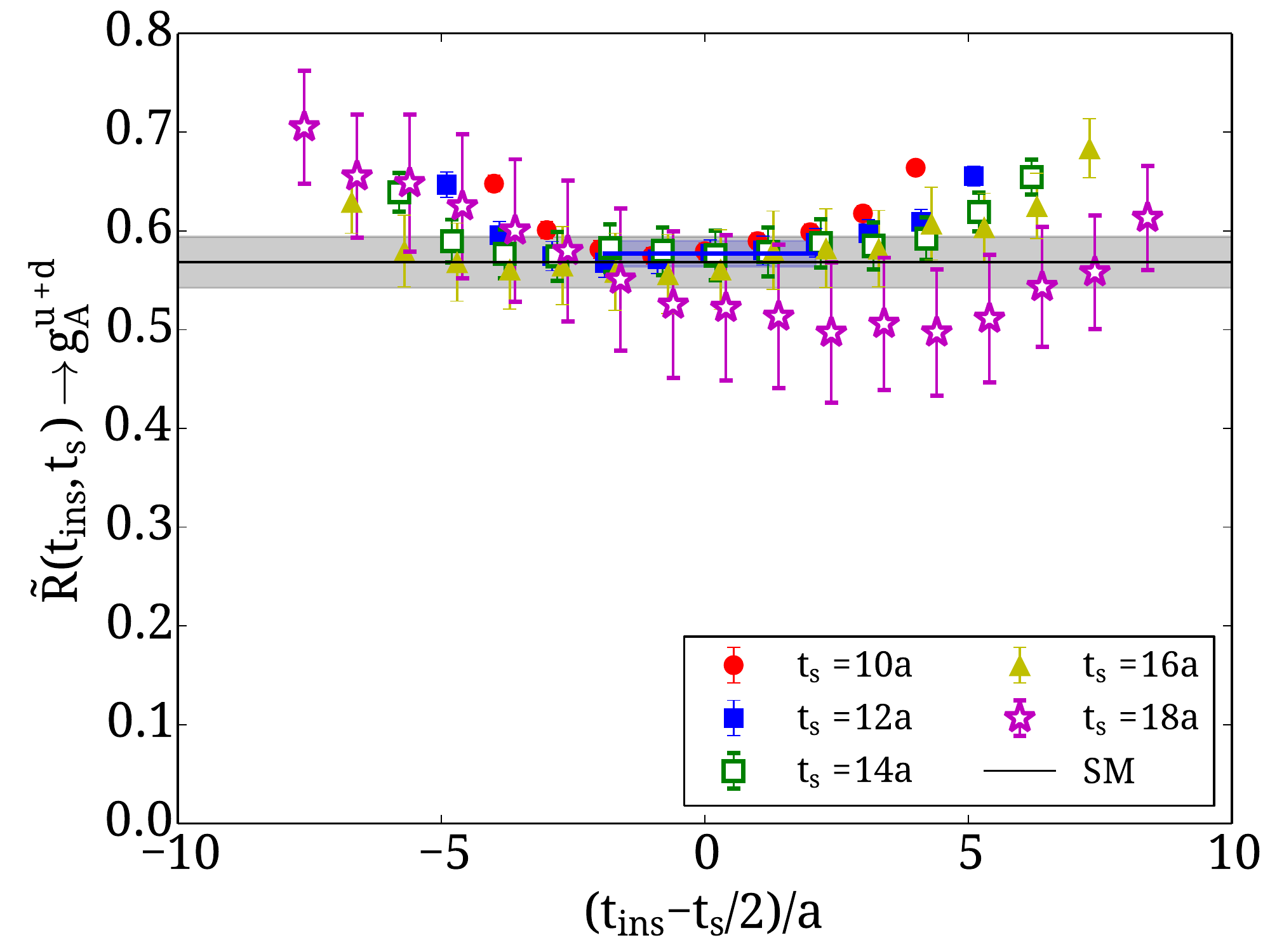}
  \caption{Connected contributions to the ratio yielding $\sigma_{\pi N}$
    (upper) and nucleon isoscalar  axial charge (lower), for various
    source-sink time separations are shown. Results obtained from a fit to a constant to the  ratio (colored band) and from a linear fit to the summed ratio (gray
    band) are also displayed.\label{zero_mom_conn_plat}}.
\end{figure}
\begin{figure}
  \includegraphics[width=1\linewidth,angle=0]{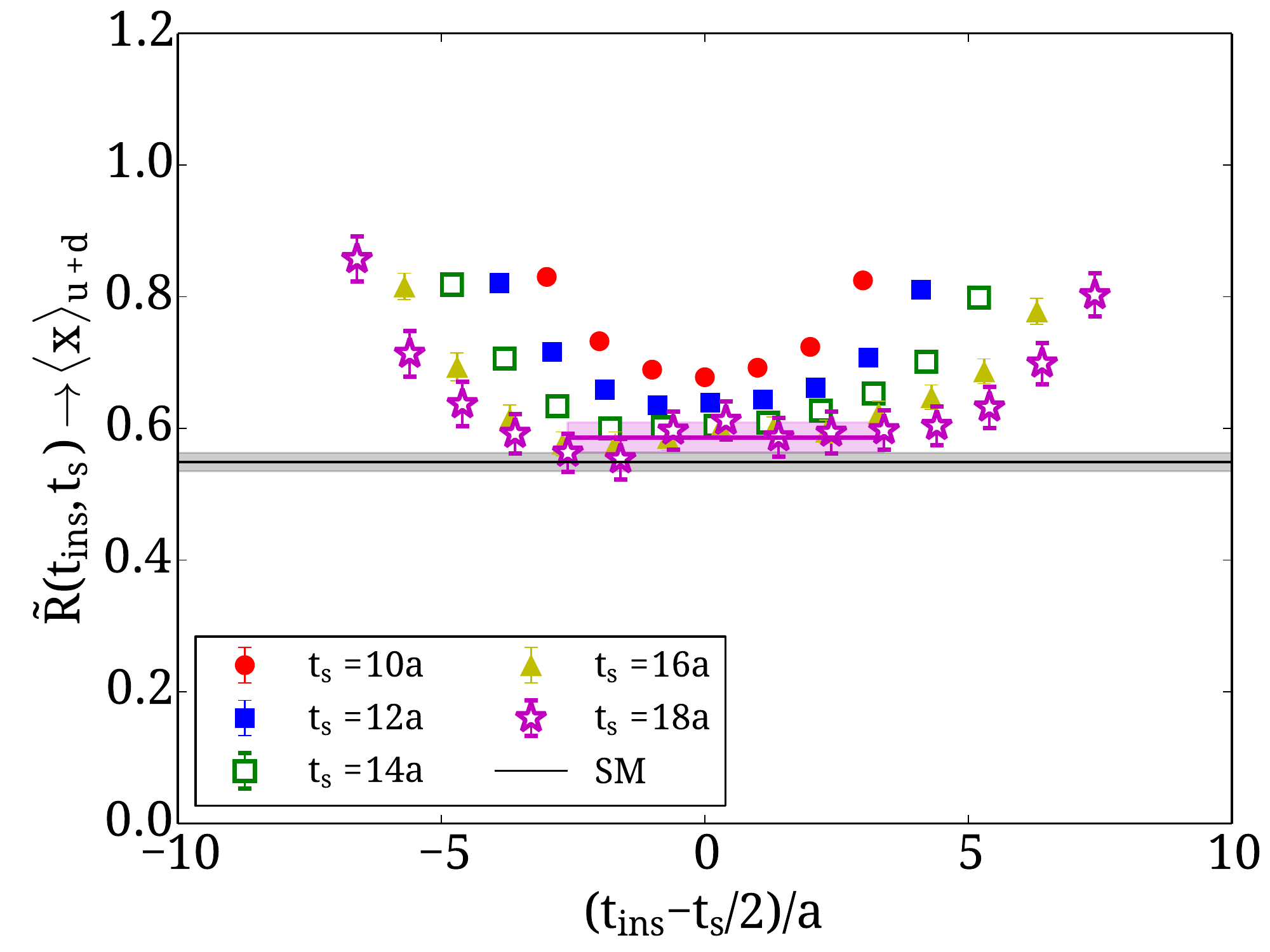}\\
  \includegraphics[width=1\linewidth,angle=0]{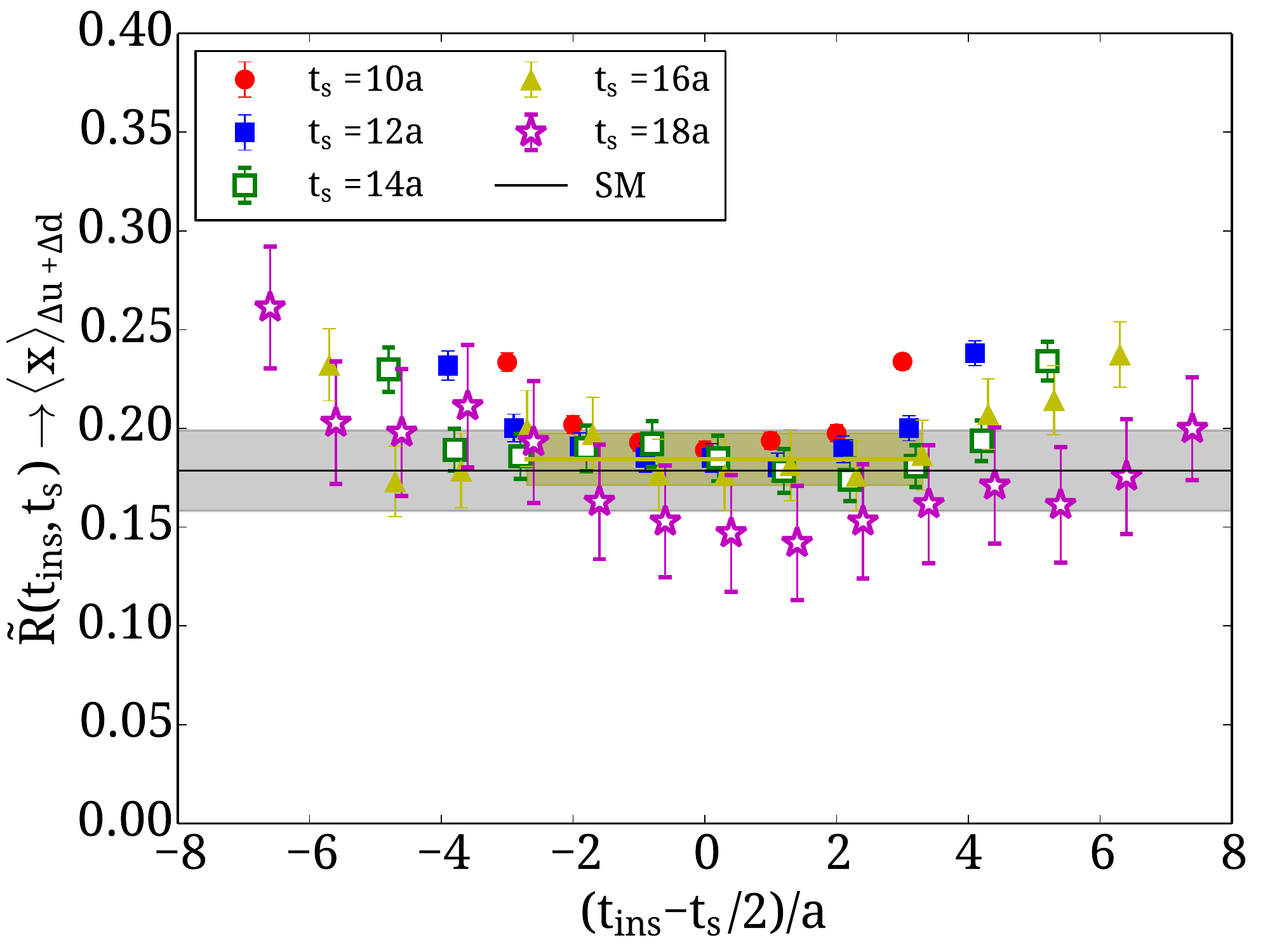}\\
  \includegraphics[width=1\linewidth,angle=0]{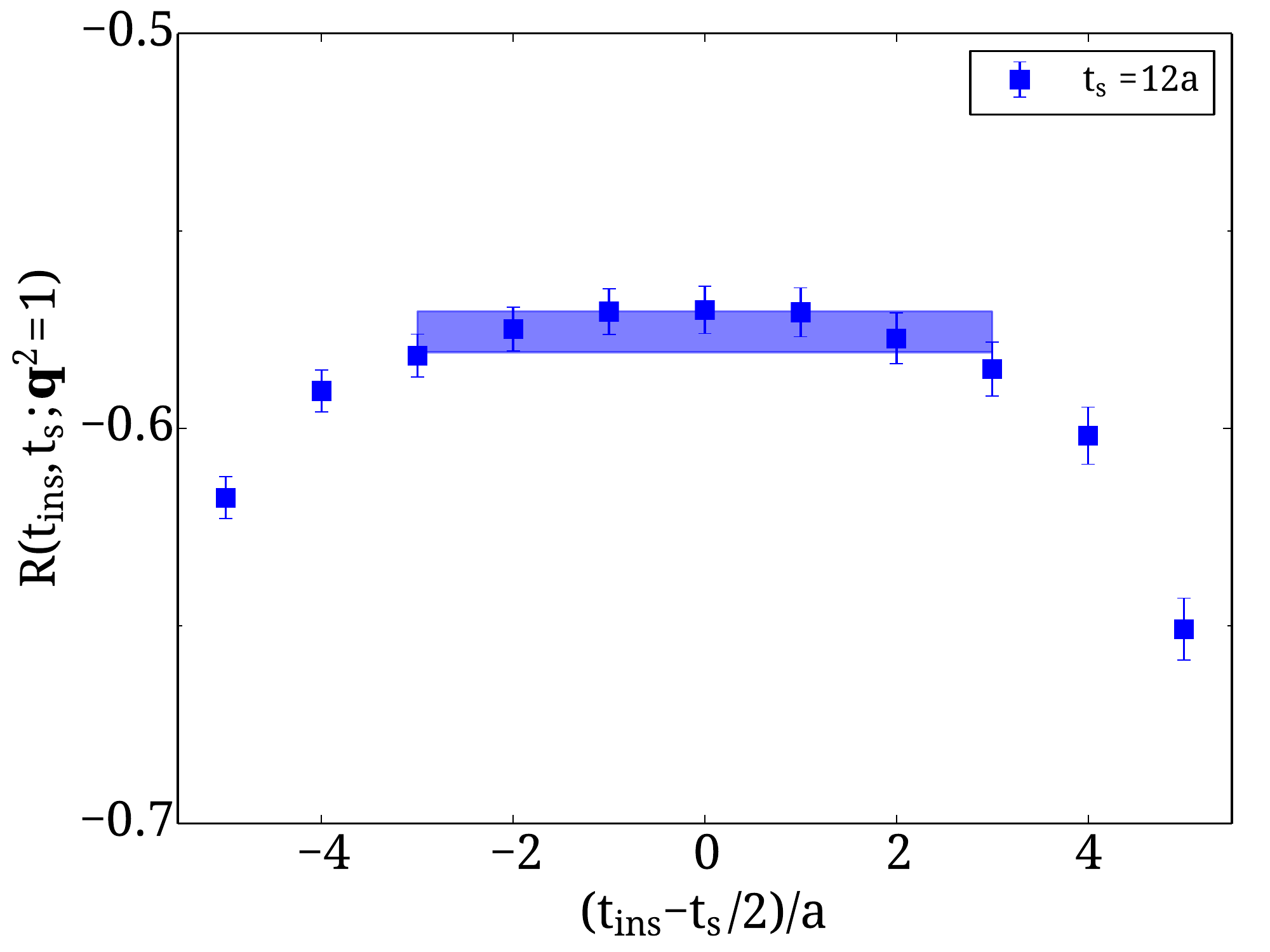}
  \caption{Connected contributions to the renormalized ratio yielding the
 isoscalar nucleon momentum
    fraction (upper), the isoscalar nucleon helicity moment (center)
    and  the axial and pseudo-scalar form
    factors $G_A(Q^2)$ and $G_p(Q^2)$ at a single unit of momentum
    (lower) are shown. For the momentum fraction and helicity, we show the results obtained from a fit to a constant to the  renormalized ratio (colored band) and from a linear fit to the summed renormalized ratio (gray
    band). \label{zero_mom_conn_one_deriv_plat}}.
\end{figure}

Finally we comment
on the issue of correlations. 
The summation and plateau methods   for various 
quantities are compared using the same set of gauge configurations 
and found to be consistent. Since these results can be  correlated,
the difference between the results of the two methods maybe underestimated.
Thus,  it is worthwhile to investigate
the  two methods using different sets of configurations.
 To perform this check  we split our ensemble into two equal sets,
which we will refer to as A-set and B-set, and redo our
analysis on these two sets separately. We show the result for the case
of the strange-quark contribution to the axial charge in
Fig.~\ref{gA_strange_parts}. As can seen,  the values computed in each set both using the plateau and summation methods  are in agreement. Furthermore, the plateau 
computed using the A-set is consistent with the summation method
computed using the B-set and vice versa.  This agreement indicates that the
consistency between the results extracted using the summation and plateau methods on the full ensemble
is not due to possible  correlations.

\section{Comparison with connected contribution}
\label{sec:connected}
The main motivation for calculating disconnected fermion loops is to
eliminate the systematic uncertainty, which arises when these are
omitted from calculations of hadronic matrix elements. For instance,
the nucleon axial charge is typically computed in the isovector
combination, where the fermion loops of the up- and down- quarks
cancel. However, if one is interested in the intrinsic spin fraction
carried by the individual quarks, one needs, in addition to the
isovector, the isoscalar combination, which involves disconnected
diagrams. Typically, in lattice QCD calculations up to now, the
disconnected contributions have been omitted. It is, therefore,
important to identify how large the contributions of disconnected
diagrams are, in order to bound the systematic error introduced
when these are neglected.
\newcommand{\red}[1]{\textcolor{red}{#1}}
\begin{table}[h!]
  \footnotesize
  \begin{center}
    \begin{tabular}{rl|rrr}
      \hline\hline
      \multicolumn{2}{c}{Observable} & connected      & disconnected & total  \\
      \hline
      \multicolumn{5}{c}{Results at zero momentum transfer ($Q^2=0$)}       \\
      $\sigma_{\pi N}$ &[MeV]     & 164.6(7.2)     & 16.6(2.4)  & 181.3(7.6) \\
      $\sigma_{s    }$ &[MeV]     &                & 21.7(3.6)  & 21.7(3.6)  \\
      $\sigma_{c    }$ &[MeV]     &                & 16(30)     & 16(30)     \\[6pt]
      $g_S^{u+d  }$    &          & 6.30(27)       & 0.639(95)  & 6.94(29)   \\
      $g_S^{s    }$    &          &                & 0.246(41)  & 0.246(41)  \\[6pt]
      $g_{A}^{u+d}$    &          & 0.576(13)      &-0.0699(89) & 0.506(15)  \\
      $g_{A}^{s  }$    &          &                &-0.0227(34) &-0.0227(34) \\[6pt]
      $g_{T}^{u+d}$    &          & 0.673(13)      &-0.0016(14) & 0.671(13)  \\[6pt]
      $\langle x \rangle_{u+d              }$&  & 0.586(22)    &    0.027(76) &    0.614(80)\\
      $\langle x \rangle_{\Delta u+\Delta d}$&  & 0.1948(51)   &   -0.058(22) &    0.136(23)\\[6pt]
      $J^{u }$                 & & 0.2781(94) &   -0.076(77) &    0.202(78)\\
      $J^{d }$                 & &-0.0029(94) &   -0.076(77) &   -0.078(78)\\
      $\Delta\Sigma^{u }/2$    & & 0.4273(50) &  -0.0174(75) &  0.4098(55)\\
      $\Delta\Sigma^{d }/2$    & &-0.1389(50) &  -0.0174(75) & -0.1564(55)\\ [6pt]
      \hline
      \multicolumn{5}{c}{Results for $\vec{q}^2=(2\pi/L)^2$ or $Q^2\simeq$0.19~GeV$^2$}\\
      $G^{u+d}_{E }$  &  & 2.2698(78)&   0.024(21)&   2.293(22)\\
      $G^{u+d}_{M }$  &  & 2.088(49) &  -0.066(75)&   2.022(89)\\
      $G^{u+d}_{A }$  &  & 0.5155(94)&  -0.0564(72) &   0.459(11) \\
      $G^{u+d}_{p }$  &  & 9.81(65)  &  -1.90(35) &   7.90(74) \\
      $B^{u+d}_{20}$  &  &-0.035(16) &  -0.33(29) &  -0.36(29) \\[6pt]
      $G^{p}_{E }$   & &   0.7453(32)&      0.0040(58)& 0.7493(47) \\
      $G^{n}_{E }$   & &   0.0113(32)&      0.0040(58)& 0.0153(47) \\
      $G^{p}_{M }$   & &   1.847(28) &     -0.011(42) & 1.836(31)  \\
      $G^{n}_{M }$   & &  -1.151(28) &     -0.011(42) &-1.162(31)  \\
      \hline
      \hline
    \end{tabular}
    \caption{ The connected and disconnected contributions to the
      various nucleon observables for the B55.32 ensemble are given in column two and three,
      whereas column four has the total contribution. The form factors
      $G_E$, $G_M$, $G_A$ and $G_p$, and generalized form factor
      $B_{20}$ are given for $\vec{q}=2\pi/L$. The disconnected
      contributions were obtained using about 150,000
      measurements.}
    \label{tab:comparison} 
  \end{center}
\end{table}

In order to assess the importance of disconnected contributions, we evaluate the connected contributions to the
isoscalar matrix elements of the operators discussed in the previous
section.  In Figs.~\ref{zero_mom_conn_plat}
and~\ref{zero_mom_conn_one_deriv_plat} we show the renormalized ratios from which the  connected part of the isoscalar
matrix elements are extracted. These results are obtained using 1200 gauge field
configurations and inverted for multiple source-sink time separations to
allow applying the summation method. We stress that, for the evaluation of the
 connected contributions unlike the case of the disconnected, to obtain
multiple source-sink time separations one needs to do a new  set of inversions
for each sink-source time separation.

The multiple source-sink time separations are computed more efficiently
by using the EigCG~\cite{Stathopoulos:2007zi,Stathopoulos:2009zz} method to deflate the
lowest eigenvalues with every new right-hand-side. For the connected
contributions shown here, we compute the sequential propagators for
eight source-sink time separations, namely from $t_s=4a$ to $t_s=18a$ for
every even time separation. In addition, the sequential propagators are
computed for both unpolarized and polarized nucleon sinks, meaning in
total 16 sequential propagators per configuration, or 16$\times$12=192
right-hand-sides are needed,
 one for each color-spin component. Our EigCG is set
up such that ten eigenvalues per right-hand-side are deflated,
stopping after a total of 24 right-hand-sides, after which the
deflated space is kept constant at 240 eigenvalues for the remaining
168 right-hand-sides. With this setup, and at this pion mass, we
observe a speedup of more than 3 times, i.e. the 192
right-hand-sides are computed for the same computational cost needed
to compute 64 right-hand-sides when not using EigCG.

The ratios yielding the connected contribution to $\sigma_{\pi N}$,
and the isoscalar $g_A$ are shown in Fig.~\ref{zero_mom_conn_plat}.
These can be compared with the corresponding ratios yielding the
disconnected contributions to $\sigma_{\pi N}$ and isoscalar $g_A$
shown in Figs.~\ref{sigmaLight} and ~\ref{gALight}, respectively.  As
can be seen, the behavior of the connected contributions is similar to
the disconnected ones, namely the sigma term shows large excited
state contamination requiring large sink-source separations whereas in
the case of $g_A^{u+d}$ the excited states are negligible even for
$t_s/a=10$. For a better comparison between connected and disconnected
contributions we collect the results extracted from the plateau method
for all nucleon observables in Table~\ref{tab:comparison}. The
disconnected contribution to the $\sigma_{\pi N}$ and isoscalar $g_A$
are found to be larger than 10\% of the connected contribution at this
quark mass. Clearly for both $\sigma_{\pi N}$ and $g_A^{u+d}$ these
are sizable effects and have to be taken into account.  The scalar
charge derives from the same matrix element as the sigma term and
therefore it also requires inclusion of disconnected contributions. For
the case of the momentum fraction, the disconnected contribution is
found to be consistent with zero as can be seen in Fig.~\ref{A20}, and
therefore we can only give an upper bound to its size to be included
in the systematic error of $\langle x\rangle_{u+d}$.  For the
polarized moment $\langle x \rangle_{\Delta u+ \Delta d}$, on the
other hand, one obtains a sizable non-zero result. Note that  the disconnected contribution
is negative decreasing the value of $\langle x \rangle_{\Delta
  u+\Delta d}$ quite substantially. The disconnected contribution to the tensor charge is essentially zero not affecting its total value.

A comment can also be made for the case of the disconnected
contributions to the nucleon form factors computed at non-zero
momentum shown in Fig.~\ref{finite_mom_plat} at a single unit of
momentum transfer squared. For the electromagnetic form-factors $G_E$
and $G_M$, we find that the disconnected contributions are consistent
with zero and with magnitude less than 1\%. With the connected
contributions at this momentum transfer being of $O(1)$, this means
that the disconnected contributions will, at most, be at the 1\% level. For
the case of the axial form factor $G_A^{u+d}$, the disconnected
contribution  is  about 10\% that of the connected  and
thus, it must be included.  In the case of the pseudo-scalar form factor
$G_p$, we find that the disconnected contribution is of similar
magnitude as the connected one and thus it is crucial in order to get
reliable results for this observable to include the disconnected part.

Having the complete set of isoscalar matrix elements with both
connected and disconnected contributions, one can combine with the
corresponding isovector matrix elements, which do not depend on
disconnected contributions, to obtain the separate quark contributions
to nucleon matrix elements. This is done in Table~\ref{tab:comparison}
for  all the various quantities considered in this work. Namely, the up- and down-quark
contributions to the nucleon spin $\Delta\Sigma^u/2$ and
$\Delta\Sigma^d/2$ are obtained by combining the isovector and
isoscalar axial charges. Including the disconnected contributions
affects the values of the intrinsic spin in particular in the case of
the d-quark.  In contrast, the values of the nucleon total spin $J^u$
and $J^d$, obtained by combining the isoscalar and isovector vector
generalized form-factors $A_{20}$ and $B_{20}$, are not affected and
the disconnected contributions only contribute an upper bound to the
error. Finally, the proton/neutron electric and magnetic form factors
$G^{p/n}_{E}$ and $G^{p/n}_{M}$ at a single unit of momentum transfer
squared, which for this lattice size and quark mass corresponds to
$Q^2\simeq0.19$~GeV$^2$, are obtained from the isovector and isoscalar
proton electric and proton magnetic form-factors assuming flavor-SU(2)
isospin symmetry between up- and down-quarks. Only the value of
$G_E^n$ is affected although, within error bars, it is still
consistent with the connected value.

\section{Conclusions}
\label{sec:conclusions}
The computation of disconnected contributions for flavor singlet
quantities has become feasible, due to the development of new
techniques to reduce the gauge and stochastic noise, and due to the
increase in computational resources. In this work, we use the
truncated solver method and the one-end trick on GPUs for the
determination of disconnected contributions to the nucleon matrix
elements. The usage of GPUs is particularly important, due to its
efficiency in the evaluation of disconnected diagrams using the TSM,
since GPUs can yield a large speedup when employing single- and
half-precision for the computation of the LP inversions and
contractions.  The calculation is performed for one ensemble of
$N_f=2+1+1$ twisted mass fermions using very high statistics. This is
necessary in order to reduce the gauge noise and obtain statistically significant  results.

The results for all observables are analyzed using both the plateau
and the summation methods. A careful analysis of excited states is
performed and we find that the methods yield results that are
compatible, as expected when excited states contributions are
negligible and identification of the fitting ranges in both methods
are well selected. Therefore, agreement of the values extracted with
the plateau and summation methods provides a good consistency check.
Since the one-end trick provides results for all sink-source
separations at no additional computational cost, such a check can be
always carried out.

Comparison of the connected to the disconnected contributions reveals
clearly that the latter are important for a number of observables
related to nucleon structure. For the sigma terms and scalar charge
the disconnected contributions amount to 10\% the total value and
thus they must be taken into account.  Similarly for the isoscalar
axial charge we find more than 10\% contributions  that must be
taken into account in the discussion of the spin carried by quarks in
the proton. The disconnected contribution reduces the value of
$\Sigma^d$ by more than 10\%, an effect that is important if we aim at a few \%
accuracy.  On the other hand, we find that the disconnected
contributions to the electromagnetic form factors at low $q^2$-values
are less than 1\% at this pion mass.  For the axial form factor $G_A$ the
disconnected contributions are sizable and persist at the level of
10\% of the value of the connected contribution even at non-zero
momentum-transfer. For $G_p$ the disconnected contribution is even larger
reaching 20\%.

In the future we plan to compute the disconnected contributions to
these quantities using simulations at physical pion mass. Such a
computation will require very large computational resources in order
to obtain results with meaningful statistical errors.

\section*{Acknowledgments}  
A. V. and M. C. are supported by funding received from the Cyprus
Research Promotion Foundation under contracts EPYAN/0506/08 and and
TECHNOLOGY/$\Theta$E$\Pi$I$\Sigma$/0311(BE)/16 respectively. K. J.  is
partly supported by RPF under contract
$\Pi$PO$\Sigma$E$\Lambda$KY$\Sigma$H/EM$\Pi$EIPO$\Sigma$/0311/16. This
research was in part supported by the Research Executive Agency of the
European Union under Grant Agreement number PITN-GA-2009-238353 (ITN
STRONGnet) and the infrastructure project INFRA-2011-1.1.20 number
283286 (HadronPhysics3), and the Cyprus Research Promotion Foundation
under contracts KY-$\Gamma$A/0310/02 and NEA
Y$\Pi$O$\Delta$OMH/$\Sigma$TPATH/0308/31 (infrastructure project
Cy-Tera, co-funded by the European Regional Development Fund and the
Republic of Cyprus through the Research Promotion
Foundation). Computational resources were provided by the Cy-Tera
machine and Prometheus (partly funded by the EU FP7 project PRACE-2IP
under grant agreement number: RI-283493) of CaSToRC, Forge at NCSA
Illinois (USA), Minotauro at BSC (Spain), and by the Jugene Blue
Gene/P machine of the J\"ulich Supercomputing Center awarded under
PRACE.  \bibliography{ref}

\end{document}